\documentclass[11pt,a4paper]{article}

\usepackage[utf8]{inputenc}
\usepackage[T1]{fontenc}
\usepackage{lmodern}
\usepackage[margin=1in]{geometry}
\usepackage{microtype}
\usepackage{setspace}
\usepackage{amsmath,amssymb,mathtools,bm} 

\usepackage{graphicx}
\usepackage{booktabs}
\usepackage{array}

\usepackage{xcolor}
\definecolor{myblue}{HTML}{1F77B4}

\usepackage[authoryear,round]{natbib}

\usepackage{algorithm}
\usepackage{algpseudocode}
\makeatletter
\renewcommand{\ALG@name}{Algorithm}
\makeatother

\usepackage{authblk}

\usepackage{hyperref}
\hypersetup{
  colorlinks=true,
  linkcolor=myblue,
  citecolor=myblue,
  urlcolor=myblue,
  pdfauthor={Bardia Panahbehagh, Mehdi Mohebbi, Amir HosseiniNasab},
  pdftitle={Intelligent n-Means Spatial Sampling}
}
\title{Intelligent n-Means Spatial Sampling}

\author[1]{Bardia Panahbehagh}
\author[2]{Mehdi Mohebbi}
\author[3]{Amir Mohammad HosseiniNasab}

\affil[1]{Faculty of Mathematical Sciences and Computer, Kharazmi University, Tehran, Iran\\
\texttt{panahbehagh@khu.ac.ir}}
\affil[2]{School of Mathematics, Statistics and Computer Science, University of Tehran, Tehran, Iran\\
\texttt{mehdi.mohebbi@ut.ac.ir}}
\affil[3]{Faculty of Science, Vrije Universiteit Amsterdam, Amsterdam, The Netherlands\\
\texttt{s.a.m.hosseininasab@student.vu.nl}}

\date{\today}

\begin{document}
\maketitle

\begin{abstract}
Well-spread samples are desirable in many disciplines because they improve estimation when target variables exhibit spatial structure. This paper introduces an integrated methodological framework for spreading samples over the population’s spatial coordinates. First, we propose a new, translation-invariant spreadness index that quantifies spatial balance with a clear interpretation. Second, we develop a clustering method that balances clusters with respect to an auxiliary variable; when the auxiliary variable is the inclusion probability, the procedure yields clusters whose totals are one, so that a single draw per cluster is, in principle, representative and produces units optimally spread along the population coordinates—an attractive feature for finite population sampling. Third, building on the graphical sampling framework of Panahbehagh (2025), we design an efficient sampling scheme that further enhances spatial balance. At its core lies an intelligent, computationally efficient search layer that adapts to the population’s spatial structure and inclusion probabilities, tailoring a design—to each specific population to maximize spread. Across diverse spatial patterns and both equal‐ and unequal‐probability regimes, this intelligent coupling consistently outperformed all rival spread‐oriented designs on dispersion metrics, while the spreadness index remained informative and the clustering step improved representativeness.
\end{abstract}



\section{Introduction}
Spreading samples is a central concern in studies involving spatially referenced data, particularly in environmental, ecological, and agricultural applications where spatial autocorrelation is prevalent. A well-spread sample improves domain coverage and thereby enhances the precision of estimators for spatially structured variables. Moreover, spatial spread often coincides with balance on auxiliary variables, which is crucial when target parameters depend nonlinearly on those covariates \citep{graf:lund:13}.

A substantial body of work has proposed designs that explicitly target spatial balance. One influential line maps a multidimensional population to a one-dimensional, neighborhood-preserving index and then applies systematic selection, as in Generalized Random Tessellation Stratified sampling (GRT; \citealp{Stev:Olse:spat:2004}). A second line induces repulsion among nearby units to discourage joint selection: the Local Pivotal Method (LPM; \citealp{gra:lun:sch:12}) iteratively updates neighboring pairs so that exactly one is retained, while Spatially Correlated Poisson sampling (SCP; \citealp{graf:11}) introduces dependence structures that reduce the likelihood of selecting adjacent units. Weakly Associated Vectors (WAV; \citealp{jauslin2020spatial}) extend this principle by iteratively perturbing the inclusion–probability vector along directions weakly associated with spatial proximity, preserving first-order inclusion probabilities while progressively driving its components to the boundary $\{0,1\}$. Partitioning and quasi-random constructions form another family, such as Halton Iterative Partitioning (HIP; \citealp{robertson2018halton}), which achieves low-discrepancy coverage through recursive subdivisions of the spatial domain. Recently, optimization-based approaches such as Dynamic Assignment Sampling (DAS; \citealp{robertson2024}) have emerged, combining geometric spread with adaptive control of sample size.

Assessing “how well-spread” a sample is typically relies on indices such as Moran \citep{moran1950notes, til:dic:esp:giu:18}, which measures spatial autocorrelation, the Voronoi index \citep{Stev:Olse:spat:2004}, and the Balanced Voronoi index \citep{prentius2024spatial}. While informative, these indices can be insensitive to certain spatial configurations, may lack directional sensitivity—e.g., distinguishing over-concentration from over-dispersion—and may not be invariant to uniform translations of the sample configuration.

Building upon this background, the present study introduces a new sampling method that serves as a competitive alternative to existing spatially balanced designs. Our framework provides a unified approach to measuring, constructing, and exploiting spatial balance under both equal- and unequal-inclusion probability settings (EP and UP). Specifically, we (i) introduce the translation-invariant density disparity index, which evaluates a sample relative to a population-specific optimal configuration and is sensitive to both under- and over-dispersion; (ii) develop a $n$-means balanced clustering scheme that forms probability-balanced, spatially compact groups—when the auxiliary variable is the inclusion probability, each cluster is in principle representative with a sample of size one, yielding units optimally spread along the population index; and (iii) design an efficient sampling algorithm—built on the graphical framework of \citet{panahbehagh2025geometric}—that enhances spatial balance while supporting design-based inference. The method incorporates an intelligent search component that examines the spatial configuration and inclusion probabilities of the population to tailor a sampling design specifically suited to its structure, thereby improving spread and representativeness without added computational burden.

Together, the three components—the density disparity index, $n$-means balanced clustering, and the graphical sampling method—form an efficient and effective framework for both measuring and achieving spatial spreadness of sample selection; moreover, when coupled with a lightweight intelligent search that dynamically reorders the design, the procedure performs population-specific optimization of spread by adapting to the observed coordinates and inclusion probabilities, thereby delivering locally tailored dispersion gains without altering first-order inclusion probabilities.

To this end, the paper is organized as follows. Section~\ref{Sec:Notation} introduces notation and design-based preliminaries for finite-population sampling. Section~\ref{Sec:Index} reviews existing indices for assessing spatial balance. Section~\ref{sec:n-means-balanced} presents $n$-means balanced clustering. Section~\ref{Sec:Measure} introduces our new measure of spatial balance, the density disparity index, and establishes its properties. In Section~\ref{Sec:SGFS}, we adapt the graphical sampling framework of \citet{panahbehagh2025geometric} to the spatial setting. Section~\ref{Sec:n-means} details the proposed sampling design, and Section~\ref{Sec:Simulation} reports the simulation results. Finally, Section~\ref{Sec:Conclusion} offers concluding remarks.

\section{Notations and Basic Concepts}\label{Sec:Notation}
 Let $U = \{1, \dots, N\}$ denote a finite population of size $N$, and let $\mathcal{S}$ be the set of all possible samples (subsets) of $U$.  
The objective is to select a sample $S \in \mathcal{S}$ with potentially unequal first-order inclusion probabilities $\bm{\pi} = \{\pi_\ell, \ell \in U\}$ and expected size $n_S$, where
\[
\sum_{\ell \in U} \pi_\ell = E(n_S).
\]
Let $\bm{p} = \{p_s = \Pr(S = s); \; s \in \mathcal{S}\}$ denote a sampling design that assigns a probability to each possible sample such that
\begin{equation}
\sum_{s \in \mathcal{S}} p_s = 1, 
\quad \text{and} \quad
\sum_{s \in \mathcal{S}, \, s \ni \ell} p_s = \pi_\ell
\quad \text{for all } \ell \in U.
\label{eqq}
\end{equation}
In this article, our focus is on designs with fixed sample size $n$, for which $E(n_S) = n$ and $\operatorname{var}(n_S) = 0$.

Consider $y_\ell$ and $x_\ell$ ($\ell \in U$) as the main and auxiliary variables, respectively.  
For any subset $A \subset U$, let $\bm{a}(A) = \{a_\ell(A), \ell \in U\}$ denote a vector of indicators where $a_\ell(A) = 1$ if $\ell \in A$ and $0$ otherwise.  
The main objective of sampling is typically to estimate parameters such as the population total,
\[
Y(A) = \sum_{\ell \in U} y_\ell a_\ell(A),
\]
which can be defined similarly for the auxiliary variable $x$.
Estimation of such parameters and their variances is based on the sampling design, the first-order inclusion probabilities, and the second-order inclusion probabilities.  
An unbiased estimator of $Y$ is the Narain–Horvitz–Thompson (NHT) estimator \citep{nar:51, hor:tho:52}:
\[
\hat{Y}(A) = \sum_{\ell \in U} \frac{y_\ell}{\pi_\ell} a_\ell(A \cap S),
\]
with variance
\begin{equation}\label{var}
\operatorname{var}(\hat{Y}(A)) = 
\sum_{\ell \in U} \sum_{\ell' \in U}
\frac{y_\ell}{\pi_\ell} \frac{y_{\ell'}}{\pi_{\ell'}} 
(\pi_{\ell \ell'} - \pi_\ell \pi_{\ell'}) 
a_\ell(A) a_{\ell'}(A),
\end{equation}
and an unbiased estimator given by
\begin{equation}\label{varest}
\widehat{\operatorname{var}}(\hat{Y}(A)) =
\sum_{\ell \in U} \sum_{\ell' \in U}
\frac{y_\ell}{\pi_\ell} \frac{y_{\ell'}}{\pi_{\ell'}} 
\frac{\pi_{\ell \ell'} - \pi_\ell \pi_{\ell'}}{\pi_{\ell \ell'}} 
a_\ell(A \cap S) a_{\ell'}(A \cap S),
\end{equation}
or a more stable estimator for fixed-size designs:
\begin{equation}\label{varest2}
\widehat{\operatorname{var}}(\hat{Y}(A) =
\frac{1}{2} \sum_{\ell \in U} \sum_{\ell' \in U}
\left( \frac{y_\ell}{\pi_\ell} - \frac{y_{\ell'}}{\pi_{\ell'}} \right)^{2}
\frac{\pi_{\ell \ell'} - \pi_\ell \pi_{\ell'}}{\pi_{\ell \ell'}} 
a_\ell(A \cap S) a_{\ell'}(A \cap S).
\end{equation}

Each population unit $\ell \in U$ is associated with a spatial coordinate denoted by $\bm{c}_\ell$.  
The spatial configuration of all population units in a $D$-dimensional space is represented by the matrix
\[
\bm{C} = \{\bm{c}_\ell, \ell \in U\}, \quad \text{where} \quad
\bm{c}_\ell = (\bm{c}_{\ell d}, \, d = 1, 2, \dots, D),
\]
and $\bm{c}_{\ell d}$ denotes the coordinate of unit $\ell$ along the $d$th dimension.  
Here, $D$ represents the dimensionality of the spatial domain under consideration.

\section{Spatial Balance Indices}\label{Sec:Index}

In this section, we discuss three indices that have been introduced to measure the spatial spread of samples: the Moran Index, the Voronoi Index, and the Balanced Voronoi Index.

\subsection{Voronoi Index (VI)}

\citet{Stev:Olse:spat:2004} introduced the Voronoi Spatial Balance Measure to assess the spatial distribution of sample units within a population.  
Let $\{U_\ell\}_{\ell \in S}$ denote the Voronoi partition of the population $U$ around the sample units $S$, such that $U_\ell$ is the set of population units closer to sample unit $\ell \in S$ than to any other.  
The VI measures how well the sample $S$ is distributed across the population $U$ and is defined as
\[
\operatorname{VI}(S) = n \sum_{\ell \in S} \left( \hat{G}_{\pi}(U_\ell) - G_{\pi}(U_\ell) \right)^2,
\]
where
\[
G_x(U^*) = F_x(U^*), \quad  
\hat{G}_x(U^*) = \frac{1}{X(U)} \sum_{\ell \in U} \frac{x_\ell}{\pi_\ell} a_\ell(U^*), 
\quad \text{and} \quad 
F_x(U^*) = \frac{X(U^*)}{X(U)}.
\]
This index reflects the deviation of the observed inclusion probabilities within each Voronoi cell from their expected values under a uniform distribution.

\subsection{Balanced Voronoi Index (BI)}
\citet{prentius2024spatial} proposed an improved Voronoi--based measure that explicitly accounts for the balance of auxiliary variables within each Voronoi region. This enhanced index is computed as a weighted sum of squared discrepancies between the design-expected and design-estimated auxiliary totals within the respective regions, thereby capturing both spatial spread and covariate balance. To balance the VI on covariates $\bm{x}$, define the scaled cellwise discrepancy $\bm{r}_{\bm{x}}(U_\ell^*)$ as the difference between the HT estimate and its design expectation (optionally scaled by a linear operator $X(U)$ to ensure commensurability of components). The index is then
\[
\operatorname{BI}(S) := \sqrt{ \frac{1}{n} \sum_{\ell \in S} \bm{r}_{\bm{x}}(U_\ell^*)^\top \mathbf{Q}^{-1} \bm{r}_{\bm{x}}(U_\ell^*) },
\]
where discrepancies are aggregated via a positive definite weighting matrix $\mathbf{Q}$. Typical choices include $\mathbf{Q}=\mathrm{Cov}(\bm{r}_{\bm{x}}(U_\ell^*))$ (design- or model-assisted) or a diagonal variance matrix, with $\mathbf{Q}=\mathbf{I}$ recovering an unweighted Euclidean aggregation.
\par\noindent The use of $\mathbf{Q}$ instead of the covariance matrix is proposed to allow the inclusion of balancing with respect to the size of the neighbourhood. A specific choice for $\mathbf{Q}$ in this context is $\mathbf{Q} = \mathbf{x}^\top \mathbf{x}$, where $\mathbf{X}$ is the Gram matrix of $\mathbf{x} = [1, x_1, \ldots, x_p]$. The resulting measure, related to the \textbf{Mahalanobis distance}, measures the squared discrepancies in the neighbourhoods around the sample units, with respect to $\mathbf{Q}$. Also, n the univariate case,
\[
r_{x}(U^*) := X(U)\big[ \widehat{G}_{x}(U^*) - G_{x}(U^*) \big],
\]
and a lower value of $\operatorname{BI}(S)$ indicates better balance of auxiliary variables across Voronoi cells.

\subsection{Moran Index (MI)}

\citet{moran1950notes} proposed a measure of spatial correlation, particularly suitable for data arranged on a grid.  
\citet{til:dic:esp:giu:18} introduced a normalized formulation using a spatial weight matrix $W$.  
Various approaches exist for defining $W$, as detailed by \citet{jauslin2020spatial} and \citet{til:dic:esp:giu:18}.  

MI quantifies spatial autocorrelation within a sample $S$, indicating whether similar values cluster or disperse spatially. It is defined as
\[
\operatorname{MI}(S) =
\frac{[\bm{a}(S) - \bar{\bm{a}}(S)]^\top W [\bm{a}(S) - \bar{\bm{a}}(S)]}
{(1^\top W 1)\,[\bm{a}(S) - \bar{\bm{a}}(S)]^\top [\bm{a}(S) - \bar{\bm{a}}(S)]},
\]
where $W = [w_{\ell \ell'}]_{N \times N}$ is the spatial weight matrix with
$w_{\ell \ell'} = 1 / e_{\ell \ell'}$ for nearby units and $w_{\ell \ell'} = 0$ otherwise, and $e_{\ell \ell'}$ denotes the distance between units $\ell$ and $\ell'$.  
Also,
\[
\bar{\bm{a}}(S) = (\bar{a}(S), \dots, \bar{a}(S)), 
\qquad
\bar{a}(S) = \frac{1}{N} \sum_{\ell \in U} a_\ell(S).
\]
A positive $\operatorname{MI}(S)$ indicates positive spatial autocorrelation (similar values cluster), while a negative value indicates negative spatial autocorrelation (dissimilar values are adjacent).

\subsection{Discussion}

While the above indices—Voronoi, Balanced Voronoi, and Moran’s—provide valuable insight into the spatial structure of a sample, each focuses on a limited aspect of spatial configuration, such as clustering, dispersion, or autocorrelation.  
They may be sensitive to edge effects, scale dependencies, or fail to reflect the uniformity of coverage across the study region.  
As shown in Section~\ref{Sec:Measure}, these limitations can lead to suboptimal assessments of spatial representativeness in certain scenarios.

To address these shortcomings, we introduce a novel index that approaches the problem of spreadness from a new perspective.  
By leveraging the cluster structure of the population prior to sampling, our proposed measure provides a more comprehensive evaluation of how evenly a sample is distributed across space, integrating both local and global dispersion patterns to yield a more robust assessment of spatial representativeness.
To define the new index and to lay the groundwork for a sampling design that achieves well-spread samples, we first introduce a clustering procedure for the population units; the next section develops this construction in detail.

\section{$n$-Means UP-balanced clustering}
\label{sec:n-means-balanced}

A central challenge in spatial and design-based sampling from finite populations is to construct clusters that achieve both geographic dispersion and precise control over inclusion probabilities.  
Classical $k$-means clustering, rooted in Lloyd’s least-squares quantization, optimizes within-cluster compactness without regard to design constraints \citep{lloyd1982least}.  
Constrained and balanced variants of $k$-means address empty clusters and enforce (near-)equal cluster sizes \citep{bradley2000constrained, demaeyer2023balanced}, yet they do not guarantee control of cluster-level sums of prescribed auxiliary totals (e.g., inclusion probabilities), which is essential for unbiased design-based inference with unequal probabilities.  
In parallel, the spatial sampling literature has developed probability designs that explicitly promote spatial spread while respecting inclusion probabilities—most notably GRTS \citep{Stev:Olse:spat:2004} and pivotal-method families that yield spatial balance and, in their doubly balanced forms, near-exact restitution of auxiliary totals \citep{gra:lun:sch:12, gra:til:13}.  
However, these are selection mechanisms rather than partitioning schemes: they select samples directly but do not produce cluster partitions in which each group meets a target auxiliary total.  
This gap motivates \emph{auxiliary-total-balanced clustering}, which partitions $U$ into spatially compact groups whose auxiliary totals match prescribed targets, thereby aligning clustering with design-based objectives and enabling one-per-cluster selection that is simultaneously well spread and probability-consistent.

In this section, we introduce a practical algorithm, termed \emph{$n$-Means UP-balanced clustering}, to partition a finite population $U$ with spatial coordinates $\bm{C}$ and inclusion probabilities $\bm{\pi} = (\pi_1, \ldots, \pi_N)$, where $\sum_{\ell=1}^N \pi_\ell = n$, into $n$ clusters $\{U_1, \ldots, U_n\}$ such that $\sum_{\ell \in U_i} \pi_\ell = 1$ for $i = 1, \ldots, n$.  
This construction enables “one-per-cluster” sampling designs that are simultaneously spatially well-spread and consistent with the specified inclusion probabilities.

The core idea is to convert the problem of balancing UP into a tractable size-balanced clustering problem via data expansion.  
For each unit $\ell$ with UP $\pi_\ell$, define an integer
\[
\texttt{count}_\ell = \max(1,~\operatorname{round}(\pi_\ell / \delta)),
\]
where $\delta > 0$ is a small \emph{split-size} parameter.  
We then construct an expanded dataset by replicating coordinate $c_\ell$ exactly $\texttt{count}_\ell$ times and record the mapping from each pseudocopy to its source unit $\ell$.

Next, we apply size-constrained $n$-means clustering (e.g., via \texttt{KMeansConstrained}\footnote{\url{https://pypi.org/project/k-means-constrained/}}) to the expanded dataset, yielding $n$ clusters of nearly equal size and cluster labels for all pseudocopies.

Aggregating the pseudocopy labels back to the original units yields, for each unit $\ell$, the fraction of its pseudocopies assigned to each cluster. Define the \emph{soft membership} matrix $M\in[0,1]^{N\times n}$ by
\[
    M[\ell, i] := \frac{\text{number of pseudocopies of unit $\ell$ assigned to cluster $i$}}{\texttt{count}_\ell}.
\]
Then $\sum_i M[\ell,i]=1$ and the cluster-attributed inclusion probabilities satisfy $\sum_i \pi_\ell M[\ell,i]=\pi_\ell$. A \emph{hard membership} may be obtained as $i^*(\ell)=\operatorname{arg\,max}_i M[\ell,i]$ (breaking ties by proximity to centroids), which induces a disjoint partition of the units.

While soft memberships preserve probabilistic mass exactly at the unit level, they may create multiple boundary units per cluster, which is undesirable for subsequent use. Hard memberships avoid fractional boundaries but can produce clusters whose total inclusion probabilities deviate from $1$. Our aim is therefore to retain the probabilistic fidelity of the soft assignments while enforcing, downstream, clusters that each carry total probability exactly equal to $1$ and have at most a single boundary.

To facilitate this, we impose a coherent global order on the preliminary clusters. Let \( \bm{o}_1, \dots, \bm{o}_n \) denote the centroids computed from the hard memberships. We obtain an ordered sequence by approximating the shortest open path through these centroids (nearest-neighbor initialization followed by a light 2-opt refinement suffices in practice). This yields an arrangement $U_{(1)}, \ldots, U_{(n)}$ in which consecutive clusters are geometrically proximate, encouraging natural joins at interfaces.

Conditioned on this cluster order, the within-cluster arrangement of units is aligned with neighboring clusters. For each ordered cluster $U_{(i)}$, we compute, for its members, their 1-nearest-neighbor distances to $U_{(i-1)}$ and $U_{(i+1)}$ (when defined), and sort $U_{(i)}$ by a simple contrast (e.g., distance to predecessor minus distance to successor). The member closest to $U_{(i-1)}$ is pinned as the ``left'' endpoint, and the member closest to $U_{(i+1)}$ as the ``right'' endpoint, so that concatenating consecutive blocks incurs minimal geometric discordance.

Concatenating the ordered blocks produces a single total order of all units. Traversing this order, we form cumulative sums of the original inclusion probabilities $\pi_\ell$ and place quota thresholds at integer levels $(1, 2, \ldots, n)$. A threshold that falls strictly between two units closes one cluster and opens the next. If a threshold falls within a unit of mass $\pi_\ell$, that unit is provisionally split fractionally between the adjacent clusters according to the portion of $\pi_\ell$ needed to satisfy the left quota. This yields crisp, interpretable borders while preserving target masses. Consequently, each cluster $U_i$ satisfies $\sum_{\ell \in U_i} \pi_\ell = 1$, producing an $n$-way partition that is UP-balanced and spatially well-organized for one-per-cluster selection.

The details are summarized in Algorithm~\ref{alg:nmeans-probbalanced-final}.

\begin{algorithm}[!htbp]
\caption{$n$-Means UP-balanced clustering}
\label{alg:nmeans-probbalanced-final}
\begin{algorithmic}[1]
\Require Coordinates $\bm{C}=(\bm{c}_1,\ldots,\bm{c}_N)$, UP $\bm{\pi}$ with $\sum_{\ell=1}^N \pi_\ell = n$, number of clusters $n$, expansion parameter $\delta$
\Ensure Centroids $\{\bm{o}_k\}_{k=1}^n$, membership matrix $M \in\mathbb{R}^{N\times n}$ (soft)

\Statex \textbf{Expand Data to Encode UP}
\For{$\ell = 1, \ldots, N$}
  \State $\texttt{count}_\ell \gets \max\!\big(1, \operatorname{round}(\pi_\ell/\delta)\big)$
\EndFor
\State Construct the expanded coordinate sequence $\bm{C}_{\text{exp}}=(\bm{c}_{e(1)},\ldots,\bm{c}_{e(N_{\text{exp}})})$ by repeating each $\bm{c}_\ell$ exactly $\texttt{count}_\ell$ times
\State Define $I_{\text{exp}}:\{1,\ldots,N_{\text{exp}}\}\to\{1,\ldots,N\}$ by $I_{\text{exp}}(u)=\ell$ iff $\bm{c}_{e(u)}=\bm{c}_\ell$ \Comment{maps each expanded index to its source unit}

\Statex \textbf{Constrained $n$-Means Clustering}
\State Apply equal-size constrained $n$-means to $\bm{C}_{\text{exp}}$ to obtain labels $L_{\text{exp}}(u)\in\{1,\ldots,n\}$ for $u=1,\ldots,N_{\text{exp}}$

\Statex \textbf{Soft Membership (per original unit)}
\State Initialize $M \gets \mathbf{0}_{N \times n}$
\For{$\ell = 1, \ldots, N$}
  \State $J_\ell \gets \{\,u \in \{1,\ldots,N_{\text{exp}}\} : I_{\text{exp}}(u)=\ell\,\}$
  \State $M[\ell,k] \gets |J_\ell|^{-1}\!\sum_{u \in J_\ell}\mathbb{I}\!\big(L_{\text{exp}}(u)=k\big)$ \quad for $k=1,\ldots,n$
\EndFor

\Statex \textbf{Preliminary Hard Labels and Centroids}
\State $z_\ell^{\text{raw}} \gets \arg\max_{k} M[\ell,k]$ \quad (ties by nearest mean in $\{\bm{c}_\ell\}$)
\State $\bm{o}_k \gets \operatorname{mean}\{\,\bm{c}_\ell : z_\ell^{\text{raw}}=k\,\}$ \quad for $k=1,\ldots,n$

\Statex \textbf{Order Clusters Along a Path}
\State Obtain an ordered sequence $(U_{(1)},\ldots,U_{(n)})$ by approximating the shortest open path through $\{\bm{o}_1,\ldots,\bm{o}_n\}$

\Statex \textbf{Within-Cluster Alignment and Total Order $\Phi$}
\For{$i=1,\ldots,n$}
  \State Sort $U_{(i)}$ by a 1-NN contrast to $U_{(i-1)}$ and $U_{(i+1)}$ (when defined), pinning endpoints nearest to neighbors
\EndFor
\State Concatenate the ordered blocks to form the total order $\Phi:\{1,\ldots,N\}\to\{1,\ldots,N\}$; $\Phi(i)$ is the $i$-th ordered unit

\Statex \textbf{Quota Split by Cumulative UP}
\State Form $sum(t)=\sum_{u\le t}\pi_{\Phi(u)}$; place thresholds at $\{1,2,\ldots,n\}$
\State If a threshold falls strictly between $\Phi(t)$ and $\Phi(t{+}1)$, cut there; if it falls within $\Phi(t)$, split $\pi_{\Phi(t)}$ fractionally between adjacent clusters

\Statex \textbf{Snap and Promote Borders}
\State Snap near-degenerate fractional shares to $\{0,1\}$ under tight tolerances; if one side receives $1$, promote $\Phi(t)$ fully to that side (no fractional border)

\Statex \textbf{Final Centroids by Hard Clustering}
\State Convert free/border allocations to per-cluster fractions; set $z_\ell \gets \arg\max_{k}$ (fraction of unit $\ell$ in cluster $k$)
\State $\bm{o}_k \gets \operatorname{mean}\{\,\bm{c}_\ell : z_\ell = k\,\}$ for $k=1,\ldots,n$

\State \Return $\{\bm{o}_k\}_{k=1}^n,~M$
\end{algorithmic}
\end{algorithm}

Conceptually, this generalizes to any auxiliary variable—such as EP—without loss of structure or interpretation, since the same balancing principle applies regardless of the weighting scheme. It ensures design consistency by forming clusters whose compositions mirror the prescribed inclusion probabilities, thereby enabling feasible, calibrated one-per-cluster selection. Using size-constrained $n$-means that preserves spatial compactness, the clusters remain geographically coherent and deliver strong spatial balance. Algorithmically, the method is practical: it recasts inclusion-probability balancing as a tractable cardinality-constrained clustering problem solvable with standard routines, and the subsequent construction of the total order is likewise fast and scalable. The resulting procedure yields soft allocations in which consecutive ordered clusters have at most one border unit and each cluster attains total probability exactly $1$. In sum, $n$-Means UP-balanced clustering provides a systematic, extensible basis for constructing spatial strata or clusters in complex surveys, maintaining spatial coherence and enabling design-based, well-spread one-per-cluster sampling.

\subsection*{Example}
\begin{figure}[!htbp] \centering \includegraphics[width=.8\linewidth]{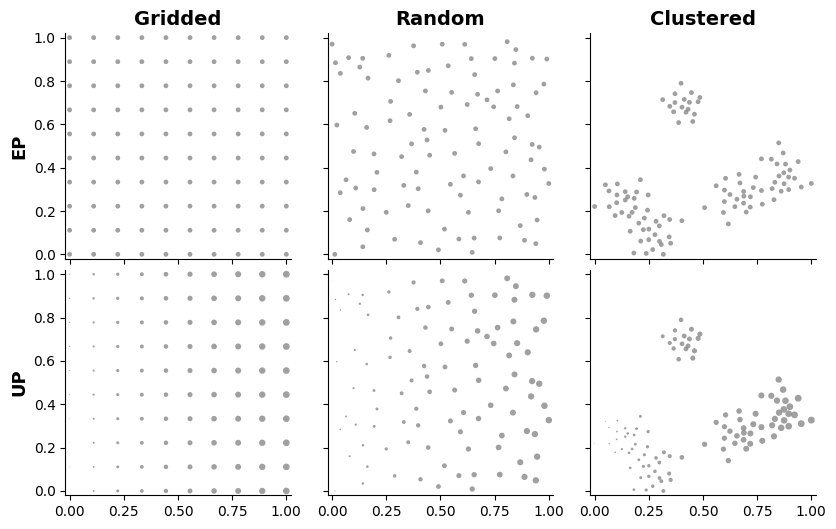} \caption{Simulated populations of size 100 under three spatial layouts: gridded (left), random (middle), and clustered (right). The top row shows equal probabilities (EP); the bottom row shows unequal probabilities (UP), where inclusion probabilities increase monotonically from left to right (larger point size indicates higher probability).} \label{fig:three_popu_original} \end{figure}

\begin{figure}[!htbp] \centering \includegraphics[width=\linewidth]{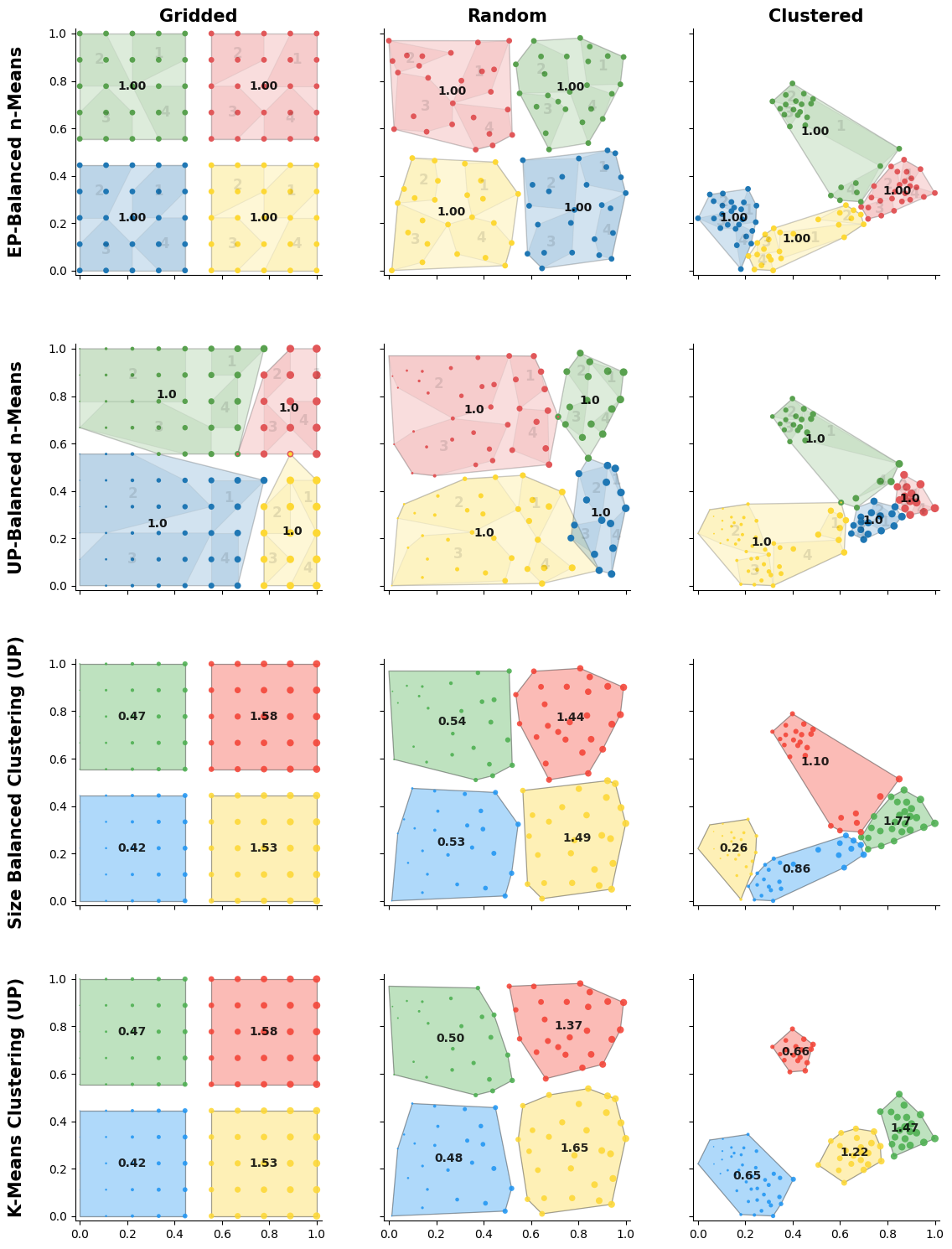} \caption{Cluster assignments for simulated populations of 100 units arranged on gridded, random, and clustered layouts. Methods compared are $n$-means balanced clustering (EP and UP), size-balanced clustering, and conventional $k$-means. In the UP case, inclusion probabilities increase from left to right. Colored polygons delineate clusters; annotated values indicate cluster inclusion-probability totals.} \label{fig:three_popu_clusters} \end{figure}

Figures~\ref{fig:three_popu_original} and~\ref{fig:three_popu_clusters} illustrate the performance of the proposed $n$-Means UP-balanced clustering compared with alternative methods under different population structures and probability regimes.

Figure~\ref{fig:three_popu_original} displays three canonical spatial populations of size $100$: gridded (left), random (middle), and clustered (right). Each layout is shown under equal probabilities (EP, top row) and unequal probabilities (UP, bottom row). In the UP setting, inclusion probabilities increase monotonically from left to right across the domain, visualized by larger point sizes. This configuration was chosen to highlight the comparative behavior of different methods.

Building on these populations, Figure~\ref{fig:three_popu_clusters} reports cluster assignments produced by four clustering strategies. The first two rows correspond to the proposed $n$-means EP- and UP-balanced clustering respectively (in general, we refer to both as $n$-means balanced clustering); the third and fourth rows display results for size-balanced clustering and conventional $k$-means. Colored polygons delineate cluster partitions, and annotated values indicate the total inclusion probability (cluster weight) in each case.

The results highlight a clear advantage of the proposed method. The $n$-means balanced clustering attains both broad spatial coverage and tight probability balance, with cluster totals summing to one across scenarios. By contrast, size-balanced and conventional $k$-means exhibit visible imbalances, particularly under heterogeneous spatial layouts or strongly unequal probabilities. This example underscores $n$-means balanced clustering as a practical tool for finite-population sampling, yielding well-spread clusters while respecting prescribed inclusion probabilities.

In addition to its clustering interpretation, the $n$-means balanced clustering can be used directly as a sampling device: each polygon in Figure~\ref{fig:three_popu_clusters} defines a region from which exactly one unit may be selected, and the balanced construction ensures that this unit is representative of the population within the polygon. In this way, the method partitions the population into spatially compact and probability-balanced clusters, so that subsequent one-per-cluster sampling yields a well-spread design that reflects both spatial structure and the specified inclusion probabilities.

\section{A New Index for Spread of Sample}\label{Sec:Measure}

In this section, we introduce a novel index for assessing the spatial spread of a sample. The core idea is that partitioning the population into clusters, each with a total inclusion probability of one, provides a meaningful representation of the population’s spatial structure for evaluating spread, and the centroids of these clusters constitute spread-optimal samples for that population. Building on this idea, we assess spread via the population’s spatial density distribution: after matching each sample unit to a cluster centroid, we translate each cluster so that its centroid coincides with its assigned unit and then compare the original and translated density surfaces; the closer these distributions, the better the sample’s spread, with near-invariance indicating a well-spread sample.

\begin{figure}[h]
    \centering
    \includegraphics[width=1.0\linewidth]{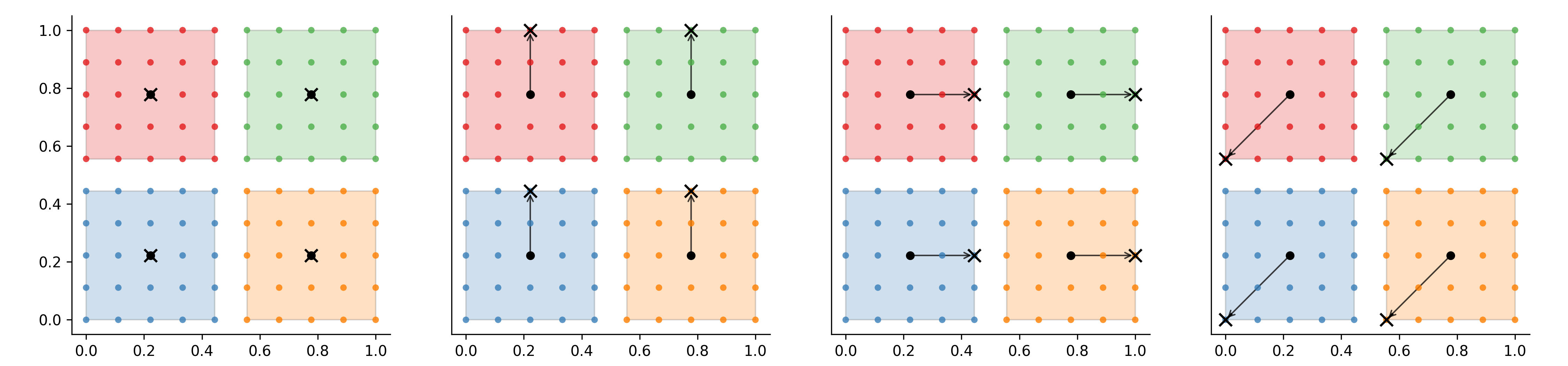}
    \caption{Spatial translation invariance of the index. In the first panel, the sample (black \(\times\)) coincides with the benchmark centroids \(O\) (black dots) within $n$-means UP-balanced clusters (colored parts) and is therefore judged optimally spread. The subsequent panels show uniform translations of this same arrangement; by translation invariance, all translated samples receive the same spread value and are likewise deemed optimal.}
    \label{fig:translation_invariant}
\end{figure}

This index exhibits several key properties that distinguish it from existing measures of spatial spread. First, it provides clear, population-specific insight into what optimal spread samples look like, as these correspond directly to the centroids obtained from $n$-Means UP-balanced clustering of the population. In contrast, for indices such as MI or Voronoi-based, the form of an optimal spread sample is neither explicit nor practically identifiable. Second, the index reveals whether a lack of spread arises from sample units being too close to one another or too far apart; such diagnostic information is absent in existing indices. Third, the index is \emph{spatially translation invariant}: the spread value of a sample remains unchanged if all its units are uniformly translated by the same vector, ensuring that absolute position does not affect the index. For example, in Figure~\ref{fig:translation_invariant}, the first panel shows a sample that coincides with the benchmark centroids and is therefore judged optimally spread by the index; uniform translations of this arrangement in the subsequent panels receive the same spread value and are likewise deemed optimal. Existing indices, by contrast, can change under such translations despite all interpoint relationships being preserved. These properties make the index a more robust and interpretable assessment of sample spread.

Since the central mechanism of this index for assessing the spread of a sample is the quantification of the disparity between the estimated density surfaces of the population before and after the translation procedure, we refer to it as the Density Disparity Index~(DI). The details of each step are presented in the remainder of this section.

\begin{figure}[h]
    \centering
    \includegraphics[width=0.75\linewidth]{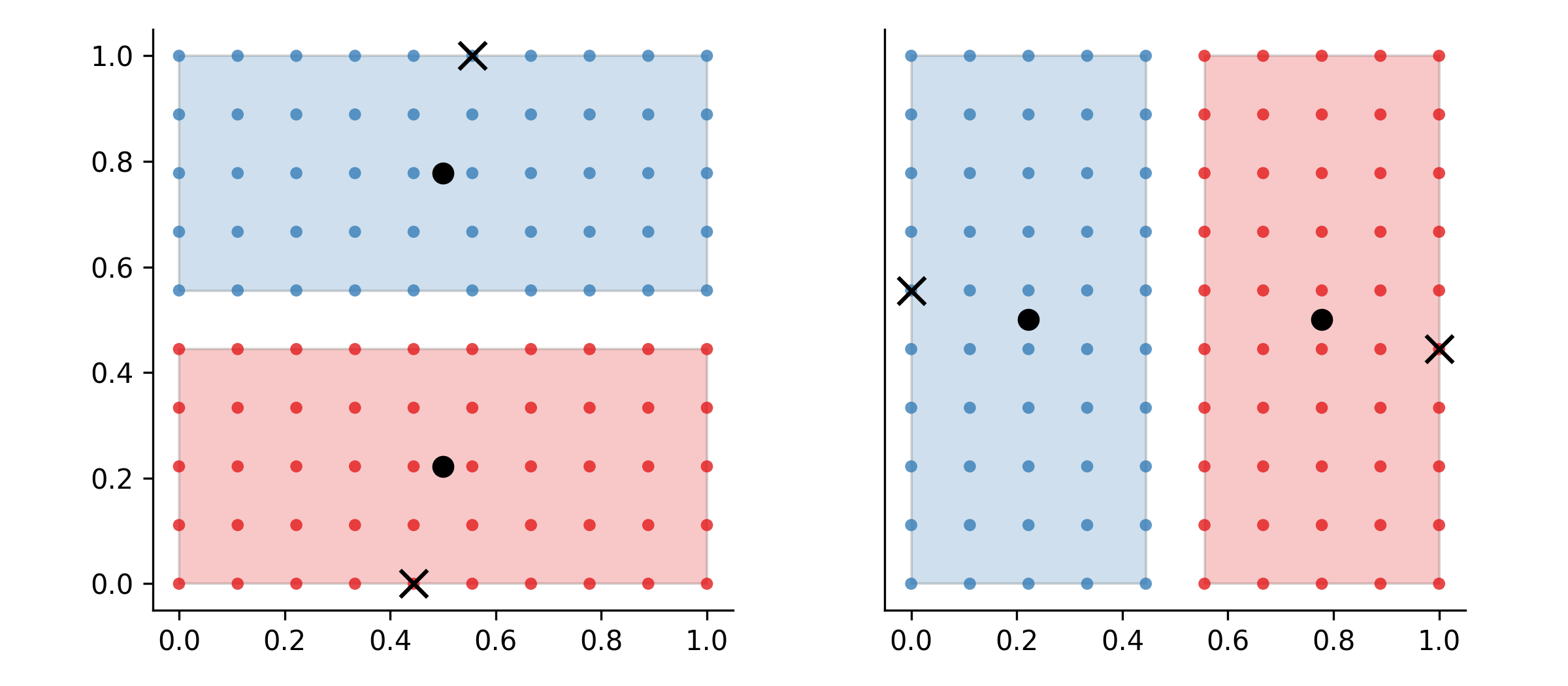}
    \caption{Sample-specific benchmark generation via \(n\)-Means UP-balanced clustering for two distinct samples of size \(n=2\) drawn from the same gridded population. The sample units of \(S\) (marked by \(\times\)) initialize the clustering. Left: A vertically arranged sample induces a horizontal partition, and the resulting benchmark centroids \(O\) (black dots) reflect this configuration. Right: A horizontally arranged sample induces a vertical partition, yielding a different set of benchmark centroids \(O\).}
    \label{fig:diff_clustering}
\end{figure}

Given a sample $S \in \mathcal{S}$, we begin by clustering the population using \(n\)-Means UP-balanced clustering, taking the units of \(S\) as the initial centroids. This partitions \(U\) into \(n\) clusters, each with a total inclusion probability of one. The centroids of these clusters, denoted \(O\), serve as the \emph{benchmark} for evaluating the spread optimality of \(S\), with the idea that the closer the arrangement of \(S\) is to that of \(O\), the more it represents an optimal spread. The arrangement of \(O\) is determined jointly by the spatial structure of the population and the arrangement of \(S\), making the benchmark both population-specific and sample-specific. This procedure allows \(O\) to adapt to each given sample. For instance, Figure~\ref{fig:diff_clustering} illustrates this with two samples for the same gridded population that have the same level of spread despite their distinct arrangements. The vertically arranged sample (left panel) induces horizontal clusters, while the horizontally arranged sample (right panel) induces vertical clusters. In each case, the sample units \(S\) are positioned near the centroids \(O\) of their respective clusters. If a single, fixed clustering were imposed—for example, the horizontal partitioning from the left panel—the second sample would be inaccurately assessed as poorly spread due to its distance from those fixed centroids. Therefore, this approach ensures that different arrangements are evaluated equitably, enabling a more appropriate comparison of their spread optimality.

Next, each unit of \(S\) is assigned to one of the centroids in \(O\) and its corresponding cluster. The arrangement of \(O\) can differ markedly from that of \(S\), particularly when \(S\) exhibits poor spatial spread. To determine an optimal one-to-one correspondence between sample units and centroids based on spatial proximity, we use the Hungarian assignment algorithm \citep{kuhn1955hungarian}. Following this assignment, the coordinates of the sample units are denoted by \(\{\boldsymbol{s}_{i} : i \in S\}\), and the coordinates of their matched centroids are denoted by \(\{\boldsymbol{o}_{i} : i \in S\}\).

With the centroid--sample assignment fixed, we introduce a clusterwise translation that recenters each population cluster at its matched sample unit in order to quantify the discrepancy between the sample arrangement and the centroid configuration. Let $U_1, \dots, U_n$ denote the clusters with centroids $\bm{o}_1, \dots, \bm{o}_n$, and let $\bm{s}_i$ be the coordinate of the sample unit matched to $\bm{o}_i$. For any unit $\ell \in U_i$ with coordinate $\bm{c}_\ell$, define
\[
T(\bm{c}_\ell) = \bm{c}_\ell + (\bm{s}_i - \bm{o}_i),
\]
which uniformly shifts all units in $U_i$ so that the cluster centroid moves from $\bm{o}_i$ to $\bm{s}_i$. Because $T$ is a rigid translation within each cluster, within-cluster pairwise distances are preserved exactly; changes arise only from the relative placement of clusters. If the sample’s spatial relationships among units closely resemble those of the centroid arrangement, then applying \(T\) to every unit has only a small effect on the relative placement of clusters. Conversely, any deviation between the sample arrangement and the centroid arrangement produces a non\mbox{-}trivial perturbation of the population’s spatial structure.

To quantify the population’s spatial structure before and after translation, we will estimate two spatial density functions via MKDE\footnote{Multivariate Kernel Density Estimation.} \citep{wand1994kernel} on the coordinates of all population units. For any location \(\boldsymbol{c}\in\mathbb{R}^2\), the MKDE estimator is
\[
    f(\boldsymbol{c}) \;=\; \frac{1}{N} \sum_{i\in S} \sum_{\ell \in U_i}
    K_{\mathbf{H}}\bigl(\boldsymbol{c} - \boldsymbol{c}_\ell\bigr),
\]
where
\begin{itemize}
    \item \(K_{\mathbf{H}}(\cdot)\) is the multivariate kernel function defined by
    \[
      K_{\mathbf{H}}(\boldsymbol{c})
      = \lvert \mathbf{H}\rvert^{-1/2}\;
        K\!\bigl(\mathbf{H}^{-1/2}\,\boldsymbol{c}\bigr),
    \]
    with \(K(\cdot)\) a symmetric multivariate density.
    \item \(\mathbf{H}\in\mathbb{R}^{2\times 2}\) is the symmetric positive‐definite bandwidth matrix, governing the amount and orientation of smoothing and selected according to Scott’s rule.
\end{itemize}
In this work we adopt the uniform kernel
\[
  K(\boldsymbol{c})
  = \frac{1}{2^2}\,\mathbf{1}\bigl\{\|\boldsymbol{c}\|_{\infty}\le1\bigr\},
\]
so that
\[
  K_{\mathbf{H}}(\boldsymbol{c})
  = \lvert \mathbf{H}\rvert^{-1/2}\,\frac{1}{2^2}\,
    \mathbf{1}\bigl\{\|\mathbf{H}^{-1/2}\boldsymbol{c}\|_{\infty}\le1\bigr\}.
\]
By applying this estimator to the original population we obtain the original spatial density surface, and by applying it to the population after cluster‐wise translations \(T(\cdot)\) we obtain the translated spatial density surface.

\begin{figure}[!htbp]
    \centering
    \includegraphics[width=1\linewidth]{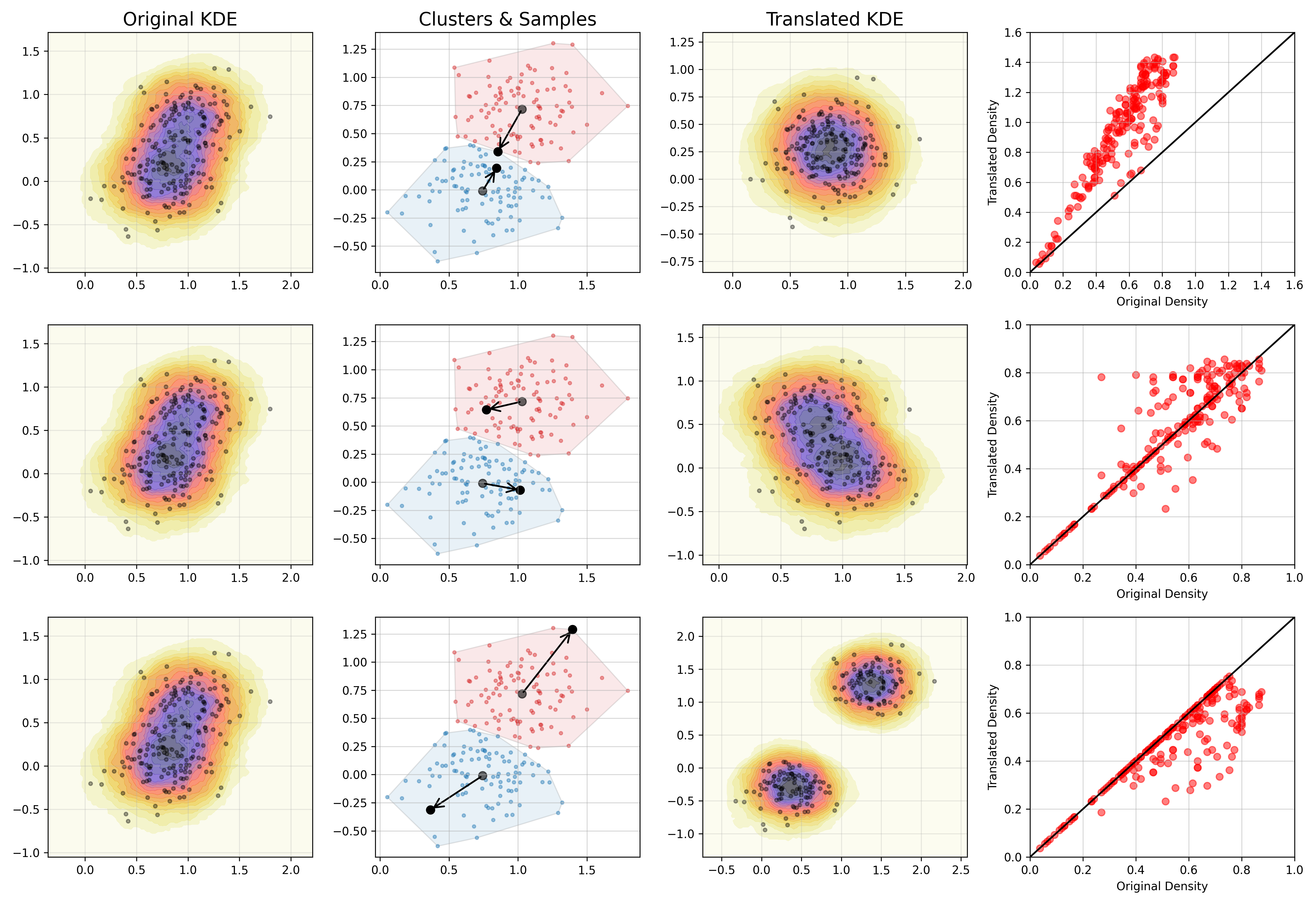}
    \caption{Illustration of cluster-wise translation and density agreement for three sample configurations in a clustered population. Columns (left to right): (i) original spatial density over the population; (ii) partition into $2$-Means UP-balanced clustering (blue/red) with sample units (bold black dots), cluster centroids (pale black dots), and arrows indicating the translation from each centroid to its assigned sample; (iii) density of the translated population after applying the cluster-wise translations; (iv) density--density comparison plotting original (horizontal) versus translated (vertical) values computed with \emph{MKDE}, where the $45^\circ$ identity line denotes equality. Rows correspond to three alternative sample arrangements (near--near, cross, and far--far), illustrating how sample arrangement influences distortion in the translated density.}
    \label{fig:kde_3_sample}

\end{figure}

Figure~\ref{fig:kde_3_sample} illustrates how cluster-wise translation affects agreement between the original and translated spatial densities. In each row, panels show (i) the original density, (ii) the $n$-Means UP-balanced clustering with sample units, centroids, and translation arrows, (iii) the translated density after applying the cluster-specific shifts, and (iv) a density--density scatter plot of translated (vertical) versus original (horizontal) values with the \(45^\circ\) identity line. The top row depicts a configuration where the sample units are too close to each other; translation causes the two clusters to overlap and locally inflate density, yielding scatter points predominantly above the identity line. The middle row shows sample units that are reasonably close to their respective centroids, but the translation arrows point in opposite directions; this introduces mild variation, so some points lie above and some below the identity line, while the majority remain near it. The bottom row presents sample units that are far apart: translation separates the clusters and locally deflates density, producing scatter points that fall below the identity line.

\begin{figure}[!htbp]
    \centering
    \includegraphics[width=0.4\linewidth]{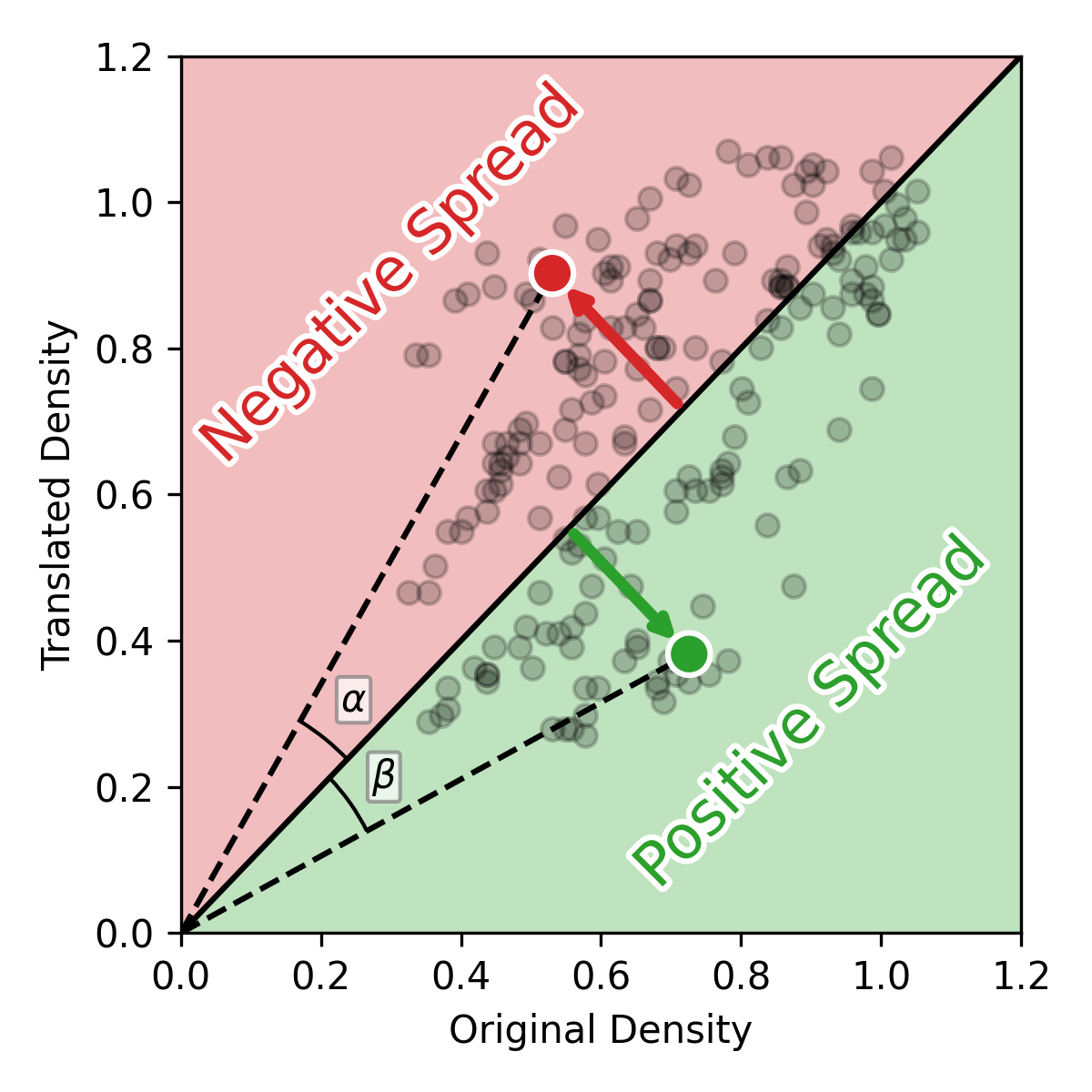}
    \caption{Angle-based contributions in the density--density scatter plot. Points show translated versus original densities with the identity line (solid) separating the \emph{negative spread} region (above; units too close) from the \emph{positive spread} region (below; units too far). For any point, the ray from the origin forms an angle with the identity line (illustrated as \(\alpha\) for the red point above and \(\beta\) for the green point below). The signed contribution to the index is proportional to the scaled value of \(-\sin\) of this angle (negative above the line, positive below), while \(1-\cos\) of the angle quantifies dissimilarity from the identity line used in the dispersion coefficient.}
    \label{fig:angles}
\end{figure}

In populations with many clusters, translation-induced effects become more intricate; some clusters converge while others diverge, placing points above or below the \(45^\circ\) identity line in the density--density scatter plot (see Figure~\ref{fig:angles}). The translated spatial structure thus emerges from inter-cluster interactions, and the distribution of points relative to the identity line succinctly conveys the global response—predominantly above indicates a net increase in local density, predominantly below indicates a net decrease, and close alignment indicates minimal structural change. Leveraging scuh plot as a diagnostic, we summarize the overall departure from optimal spread by the net deviation of points from the identity line: a net deviation of zero corresponds to a perfectly spread sample—or to an arrangement in which local contractions and expansions (some units closer, others farther) offset to yield a balanced outcome; its magnitude quantifies the loss of spread quality; and its sign reveals the mode of failure—positive when units are on average too close (inflated translated density) and negative when they are too far apart (deflated translated density). This signed characterization captures both the extent and the nature of deviation from optimal spread, offering information unavailable from conventional indexes.

To formalize the notion of \emph{Net Deviation}, we examine each point of the density–density scatter plot at the level of individual population units. For a unit \(\ell \in U\) with location \(\bm{c}_\ell\), define
\[
\mathbf{f}_\ell \;=\;
\begin{pmatrix}
f\!\bigl(\bm{c}_\ell\bigr)\\[4pt]
f\!\bigl(T(\bm{c}_\ell)\bigr)
\end{pmatrix},
\qquad
\mathbf{1} \;=\;
\begin{pmatrix}
1\\[2pt]
1
\end{pmatrix}.
\]
Let \(\alpha_\ell\) be the oriented angle from \(\mathbf{1}\) to \(\mathbf{f}_\ell\). A direct calculation yields
\[
\sin\alpha_\ell
=\frac{\|\mathbf{1}\times \mathbf{f}_\ell\|}{\|\mathbf{1}\|\,\|\mathbf{f}_\ell\|}
=\frac{\,f\!\bigl(T(\bm{c}_\ell)\bigr)-f\!\bigl(\bm{c}_\ell)\,}{\sqrt{2}\,\sqrt{f\!\bigl(T(\bm{c}_\ell)\bigr)^2 + f\!\bigl(\bm{c}_\ell\bigr)^2}},
\]
\[
\cos\alpha_\ell
=\frac{\mathbf{1}\cdot \mathbf{f}_\ell}{\|\mathbf{1}\|\,\|\mathbf{f}_\ell\|}
=\frac{\,f\!\bigl(T(\bm{c}_\ell)\bigr)+f\!\bigl(\bm{c}_\ell)\,}{\sqrt{2}\,\sqrt{f\!\bigl(T(\bm{c}_\ell)\bigr)^2 + f\!\bigl(\bm{c}_\ell\bigr)^2}}.
\]
Here, \(\sin\alpha_\ell\) is a signed angular deviation of the point \(\mathbf{f}_\ell\) from the identity line. To align signs with our diagnostic convention (negative for local inflation after translation and positive for local deflation), we use \(-\sin\alpha_\ell\), hence values \(<0\) indicate local inflation (points above the identity line), values \(>0\) indicate local deflation (points below the line), and \(0\) indicates local agreement. In order to quantify the angular dispersion (dissimilarity) of points from the identity line, we will use \(1-\cos\alpha_\ell\); the factor \(1-\cos\alpha_\ell\) will serve as a natural weight when aggregating pointwise deviations into a global signed measure.

Since \(\alpha_\ell\) typically takes small values, we apply the following scaling to increase sensitivity and capture finer differences:
\[
\operatorname{scale}(\beta;\gamma)\;=\;
\begin{cases}
1, & \frac{\beta}{\gamma} \ge 1,\\[2pt]
-1, & \frac{\beta}{\gamma} \le -1,\\[2pt]
\frac{\beta}{\gamma}, & \text{otherwise},
\end{cases}
\]
where \(\gamma>0\). Under this transformation, values with \(|\beta|\ge\gamma\) are saturated to \(\pm 1\), while those with \(|\beta|<\gamma\) are mapped linearly from \((-\gamma,\gamma)\) to \((-1,1)\).

Figure~\ref{fig:angles} provides an angle-based visualization of these definitions. Each point corresponds to a population unit with coordinates \(\bigl(f(\bm{c}_\ell),\,f(T(\bm{c}_\ell))\bigr)\); the solid \(45^\circ\) line is the identity. Points above the line indicate \emph{negative spread} (local inflation after translation), while points below indicate \emph{positive spread} (local deflation). For any point, the ray from the origin forms an oriented angle \(\alpha_\ell\) with the identity line. The signed contribution entering the net component is given by the scaled value of \(-\sin\alpha_\ell\) (negative above, positive below), and the dispersion contribution is measured by \(1-\cos\alpha_\ell\), which quantifies angular dissimilarity from the identity line.

The net deviation \(D_{\text{net}}\) is defined as the average of the scaled negative sin over all population units:
\[
D_{\text{net}}
\;=\;\frac{1}{N}\sum_{\ell\in U}\operatorname{scale}\!\left(-\sin\alpha_\ell;\,\sin(\tfrac{\pi}{8})\right).
\]

Equivalently, \(D_{\text{net}}\) decomposes into the contributions of positive and negative deviations,
\[
D_{+}
\;=\;\sum_{\ell:\,-\sin\alpha_\ell-\geq 0}\operatorname{scale}\!\left(-\sin\alpha_\ell;\,\tfrac{\pi}{8}\right),
\qquad
D_{-}
\;=\;\sum_{\ell:\,-\sin\alpha_\ell<0}\operatorname{scale}\!\left(-\sin\alpha_\ell;\,\tfrac{\pi}{8}\right),
\]
\[
D_{\text{net}}
\;=\;\frac{D_{+}+D_{-}}{N},
\]
where \(D_{+}\) captures \emph{positive spread}—the component arising when sample units are, on average, too far apart (points below the identity line)—and \(D_{-}\) captures \emph{negative spread}—the component arising when sample units are, on average, too close (points above the line).

Thus \(D_{\text{net}}\approx 0\) in two situations: (i) points lie predominantly on the identity line, indicating near-equality of original and translated densities (sample units effectively coincide with the UP-Balanced centroids, i.e., optimal spread); or (ii) nonzero positive and negative deviations are present but offset each other, reflecting an arrangement with some units closer and others farther such that their effects balance.

The second case has a useful interpretation: it identifies a form of well-spread arrangement arising from \emph{balanced imperfections} (some units too close, others too far). However, to avoid awarding near-perfect values when \(|D_{+}|\) and \(|D_{-}|\) are both large but offset, we introduce a \emph{Dispersion Coefficient} that measures the overall angular dissimilarity from the identity line, irrespective of sign, as
\[
\eta
\;=\;\frac{1}{N}\sum_{\ell\in U}\operatorname{scale}\!\left(1-\cos\alpha_\ell;\,1-\cos(\tfrac{\pi}{8})\right),
\]
which is small when the translated and original densities closely align pointwise and increases as the scatter departs from the identity line.

For a given sample \(S \in \mathcal{S}\) and population \(U\), combining the net deviation with the Dispersion Coefficient while preserving the \([-1,1]\) range yields the DI:
\[
\text{DI}(S)
\;=\;
D_{\text{net}}
\;+\;
\bigl(\operatorname{sign}(D_{\text{net}})-D_{\text{net}}\bigr)\,\eta,
\]
which interpolates between \(D_{\text{net}}\) (when \(\eta=0\)) and \(\operatorname{sign}(D_{\text{net}})\) (when \(\eta=1\)).

Algorithm~\ref{alg:ddi} summarizes the full computation of the DI: starting from a \(n\)-Means UP-balanced clustering partition initialized at the sample, it assigns centroids via the Hungarian algorithm, applies cluster-wise translations, estimates original and translated densities, computes pointwise angles, and aggregates scaled \(-\sin\alpha_\ell\) and \(1-\cos\alpha_\ell\) into the signed component \(D_{\text{net}}\) and the dispersion coefficient \(\eta\), respectively, to yield \(\text{DI}\in[-1,1]\).

\begin{algorithm}[!htbp]
\caption{Density Disparity Index (DI)}
\label{alg:ddi}
\begin{algorithmic}[1]
\Require Population \(U=\{1,\dots,N\}\) with coordinates \(\{\bm{c}_\ell\in\mathbb{R}^2\}_{\ell\in U}\); inclusion probabilities \(\bm{\pi}=(\pi_\ell)_{\ell\in U}\) with \(\sum_{\ell}\pi_\ell=n\); sample \(S=\{i_1,\dots,i_n\}\subset U\); kernel density estimator \(\text{MKDE}(\cdot)\); scaling function \(\operatorname{scale}(\beta;\gamma)\).

\Ensure Density Disparity Index \(\text{DI}\in[-1,1]\).
\vspace{3pt}
\Statex \textbf{Clustering and benchmark centroids}
\State Run \textbf{\(n\)-Means UP-balanced clustering} on \(U\) with initialization by the sample units \(S\).\label{line:upbkm}
\State Let \(\{U_i\}_{i=1}^n\) be the clusters with \(\sum_{\ell\in U_i}\pi_\ell=1\).\label{line:clusters}
\State Denote the set of centroids by \(O=\{\bm{o}_1,\dots,\bm{o}_n\}\) (benchmark).
\vspace{3pt}
\Statex \textbf{Assignment and translation}
\State Compute a one-to-one assignment \(\sigma:\{1,\dots,n\}\!\to\!\{1,\dots,n\}\) between sample units \(\{\bm{s}_{i}\}\) and centroids \(\{\bm{o}_i\}\) by the Hungarian algorithm.\label{line:hungarian}
\State Define the cluster-wise translation \(T(\bm{c}_\ell)=\bm{c}_\ell+\bm{s}_{i}-\bm{o}_i\) for all \(\ell\in U_i\).\label{line:T}
\vspace{3pt}
\Statex \textbf{Density estimation}
\State Estimate the spatial density on original and translated locations:
\[
f(\bm{c}_\ell)\;\gets\;\textsc{MKDE}\bigl(\bm{c}_\ell;\{\bm{c}_m\}_{m\in U}\bigr),\qquad
f\bigl(T(\bm{c}_\ell)\bigr)\;\gets\;\textsc{MKDE}\bigl(T(\bm{c}_\ell);\{T(\bm{c}_m)\}_{m\in U}\bigr).
\]\label{line:mkde}
\vspace{3pt}
\Statex \textbf{Pointwise angles}
\State For each \(\ell\in U\), compute
\[
\sin\alpha_\ell=\frac{f(T(\bm{c}_\ell))-f(\bm{c}_\ell)}{\sqrt{2}\sqrt{f(T(\bm{c}_\ell))^2+f(\bm{c}_\ell)^2}},\qquad
\cos\alpha_\ell=\frac{f(T(\bm{c}_\ell))+f(\bm{c}_\ell)}{\sqrt{2}\sqrt{f(T(\bm{c}_\ell))^2+f(\bm{c}_\ell)^2}}.
\]\label{line:angles}
\vspace{3pt}
\Statex \textbf{Aggregation}
\State Net deviation: \(\displaystyle D_{\text{net}} \gets \frac{1}{N}\sum_{\ell\in U}\operatorname{scale}\!\left(-\sin\alpha_\ell;\,\tfrac{\pi}{8}\right)\).\label{line:nd}
\State Dispersion Coefficient: \(\displaystyle \eta \gets \frac{1}{N}\sum_{\ell\in U}\operatorname{scale}\!\left(1-\cos\alpha_\ell;\,1-\cos(\tfrac{\pi}{8})\right)\).\label{line:eta}
\vspace{3pt}
\Statex \textbf{Index}
\State Density Disparity Index:
\[
\text{DI}\;\gets\; D_{\text{net}}+\bigl(\operatorname{sign}(D_{\text{net}})-D_{\text{net}}\bigr)\,\eta.
\]\label{line:ddi}
\State \Return \(\text{DI}\).
\end{algorithmic}
\end{algorithm}

\subsection*{Example} 
\begin{figure}[!htbp] \centering \includegraphics[width=130mm]{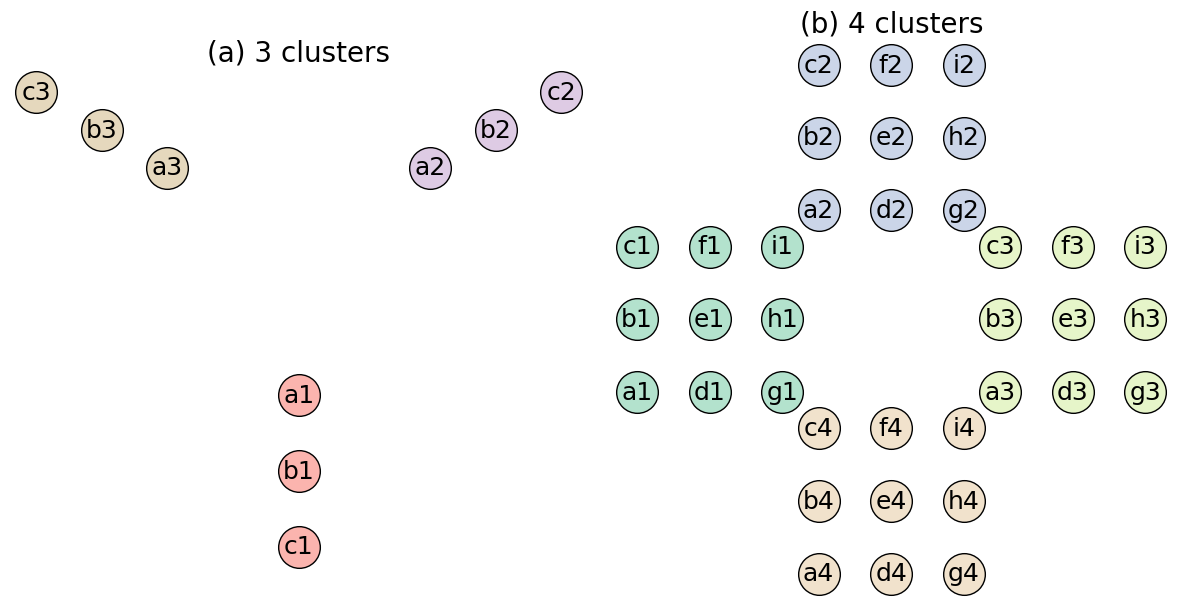} \caption{Illustrative population structures used to compare classical spread indices with the proposed Density Disparity Index (DI). Panel~(a): three rotated clusters placed far apart. Panel~(b): four uniformly arranged clusters in close proximity. These contrasting cases highlight situations where MI, the VI, or the BI may provide indistinguishable or misleading results, whereas DI adapts to the underlying population geometry and more faithfully reflects differences in spatial spread.} \label{fig:Four_Three} \end{figure}

To clarify the limitations of existing spatial balance indices and the advantages of the proposed DI, we consider two simplified population settings shown in Figure~\ref{fig:Four_Three}. In panel~(a), the population forms three well-separated clusters placed at large mutual distances. In panel~(b), the population forms four compact clusters arranged symmetrically in close proximity.

The numerical results in Table~\ref{tab:Four_Three} detail the limitations of classical indices and the strength of DI.

For the three-cluster setting, panel~(a), both the VI and the MI fail to discriminate among samples: all are evaluated identically as “perfect” ($\mathrm{VI}=0$ and $\mathrm{MI}=-1$), despite clear structural differences. The BI is more informative but is unsigned; it cannot indicate whether a sample is overly concentrated or overly dispersed and therefore does not distinguish extreme cases such as \texttt{aaa}, \texttt{aac}, and \texttt{ccc}. By contrast, DI produces a meaningful ordering: \texttt{aaa} is least well spread ($-0.84$), \texttt{aac} is intermediate ($-0.37$), and \texttt{ccc} is most spread ($0.74$). The range (from $-0.84$ to $0.74$) shows that DI captures gradations of spread rather than collapsing configurations into a single class as VI and MI do here.

For the four-cluster setting, panel~(b), the first three samples (\texttt{e1 e2 e3 e4}, \texttt{a1 a2 a3 a4}, \texttt{d1 d2 d3 d4}) are simple translations of the centroid configuration. DI correctly treats all as well spread ($0.00$), while other indices are ambiguous. The fourth sample (\texttt{f1 h2 d3 b4}) is essentially a rotation, which DI rates near balance ($0.11$). The fifth and sixth samples (\texttt{b1 f2 h3 d4} and \texttt{h1 d2 b3 f4}) move one unit far from, or very close to, another cluster; DI reflects this asymmetry in both sign and magnitude ($0.78$ versus $-0.71$). The seventh sample (\texttt{a1 d1 b1 e1}) exhibits extreme proximity, yielding the lowest DI ($-0.86$) and correctly flagging the worst imbalance. Again, BI and VI fail to discriminate many of these configurations. MI detects clustering in the extreme case (\texttt{a1 d1 b1 e1}, $\mathrm{MI}=0.61$) but is less informative in intermediate cases: \texttt{b1 f2 h3 d4} (overspread) and \texttt{h1 d2 b3 f4} (overconcentrated) both receive middling values (about $-0.50$ to $-0.69$), indicating that MI lacks the directional sensitivity that DI provides.

Taken together, these examples show that DI not only separates dense from spread samples but also provides a signed, directionally sensitive measure indicating whether a sample is too compact or too dispersed. By comparison, BI is unsigned, VI is insensitive to global configuration, and MI can be misleading in clustered layouts. DI therefore offers a richer and more interpretable diagnostic of spatial spread, responsive to both global separation and local concentration.

\begin{table}[ht]
\centering
\caption{Index values for example samples in Figure \ref{fig:Four_Three}. Columns report Density Disparity Index (DI), Balanced-Voronoi (BI), Voronoi (VI), and Moran (MI). Left block corresponds to subplot (a) with three clusters; right block to subplot (b) with four clusters.}\label{tab:Four_Three}
\resizebox{\textwidth}{!}{%
\begin{tabular}{ccccrrrr @{\hspace{2em}} cccc rrrr}
\toprule
\multicolumn{8}{c}{\textbf{(a) 3 clusters}} & \multicolumn{8}{c}{\textbf{(b) 4 clusters}} \\
\cmidrule(lr){1-8}\cmidrule(l){9-16}
\multicolumn{4}{c}{Sample} & DI & BI & VI & MI &
\multicolumn{4}{c}{Sample} & DI & BI & VI & MI \\
\midrule
a & a & a &   & -0.84 & 0.26 & 0.00 & -1.00 & e1 & e2 & e3 & e4 &  0.00 & 0.00 & 0.00 & -1.00 \\
a & a & b &   & -0.69 & 0.21 & 0.00 & -1.00 & a1 & a2 & a3 & a4 &  0.00 & 0.43 & 0.11 & -0.59 \\
a & a & c &   & -0.37 & 0.26 & 0.00 & -1.00 & d1 & d2 & d3 & d4 &  0.00 & 0.28 & 0.02 & -0.83 \\
a & b & b &   & -0.41 & 0.15 & 0.00 & -1.00 & f1 & h2 & d3 & b4 &  0.11 & 0.19 & 0.00 & -0.75 \\
a & b & c &   & -0.12 & 0.21 & 0.00 & -1.00 & b1 & f2 & h3 & d4 &  0.78 & 0.26 & 0.00 & -0.50 \\
a & c & c &   &  0.21 & 0.26 & 0.00 & -1.00 & h1 & d2 & b3 & f4 & -0.71 & 0.26 & 0.00 & -0.69 \\
b & b & b &   &  0.00 & 0.00 & 0.00 & -1.00 & a1 & d1 & b1 & e1 & -0.86 & 1.01 & 0.73 &  0.61 \\
b & b & c &   &  0.30 & 0.15 & 0.00 & -1.00 & a1 & c1 & g1 & i1 & -0.70 & 0.85 & 0.58 &  0.38 \\
b & c & c &   &  0.56 & 0.21 & 0.00 & -1.00 & c1 & c2 & g3 & g4 &  0.56 & 0.36 & 0.00 & -0.40 \\
c & c & c &   &  0.74 & 0.26 & 0.00 & -1.00 & i1 & a2 & a3 & i4 & -0.51 & 0.36 & 0.00 & -0.05 \\
\bottomrule
\end{tabular}%
}
\end{table}

\section{Graphical Sampling for Spatial Populations}\label{Sec:SGFS}
\cite{panahbehagh2025geometric} introduced a graphical framework for finite-population sampling (GFS) that affords intuitive control over sampling designs. The original development treated the one-dimensional case, where units are arranged along a single axis without explicit spatial coordinates, yet it accommodates a broad class of unequal-probability designs ranging from Madow’s systematic sampling \citep{mad:49} to maximum-entropy sampling \citep{til:06}. In this paper, we extend the algorithm to populations with $D$-dimensional coordinates, with particular emphasis on the spatial case $D=2$. Throughout, we assume a fixed-size design with $\sum_{\ell=1}^N \pi_\ell = n \in \mathbb{N}$. The resulting methodology is detailed in Algorithm~\ref{AlGFS2D}.

\begin{algorithm}[!htbp] \caption{Fixed-Size Spatial GFS for a Population with $D$-Dimensional Coordinates}\label{AlGFS2D} \begin{algorithmic}[1] 
\State Construct a $(D{+}1)$-dimensional system: represent the spatial coordinates of each population unit on the first $D$ axes, and use the $(D{+}1)$-th axis to display its inclusion probability in $(0,1]$. 
\State Order the population units (arbitrarily or at random) to obtain a sequence $\bm{c}_{1}, \bm{c}_{2}, \ldots, \bm{c}_{N}$, where each $\bm{c}_\ell \in \mathbb{R}^D$ denotes a point in the coordinate space. 
\State Initialize the first bar at $\bm{c}_{1}$ to cover $[0,\pi_{1}]$ on the $(d{+}1)$-th axis; set $b_{1} \gets \pi_{1}$. \For{$\ell = 2$ to $N$} \If{$b_{\ell-1} + \pi_{\ell} \leq 1$} 
\State Place the $\ell$-th bar at $\bm{c}_{\ell}$ covering $(b_{\ell-1},, b_{\ell-1}{+}\pi_{\ell}]$; set $b_{\ell} \gets b_{\ell-1}{+}\pi_{\ell}$. \Else \State Place the $\ell$-th bar at $\bm{c}_{\ell}$ covering the wrapped interval $(b_{\ell-1},1] \cup [0,, b_{\ell-1}{+}\pi_{\ell}{-}1]$;  \State set $b_{\ell} \gets b_{\ell-1}{+}\pi_{\ell}{-}1$. \EndIf \EndFor \State Draw $r \sim U(0,1)$. The final sample consists of all units whose assigned bar contains $r$ along the $(d{+}1)$-th axis. \end{algorithmic} \end{algorithm}

\begin{figure}
 \centering
\includegraphics[width=150mm]{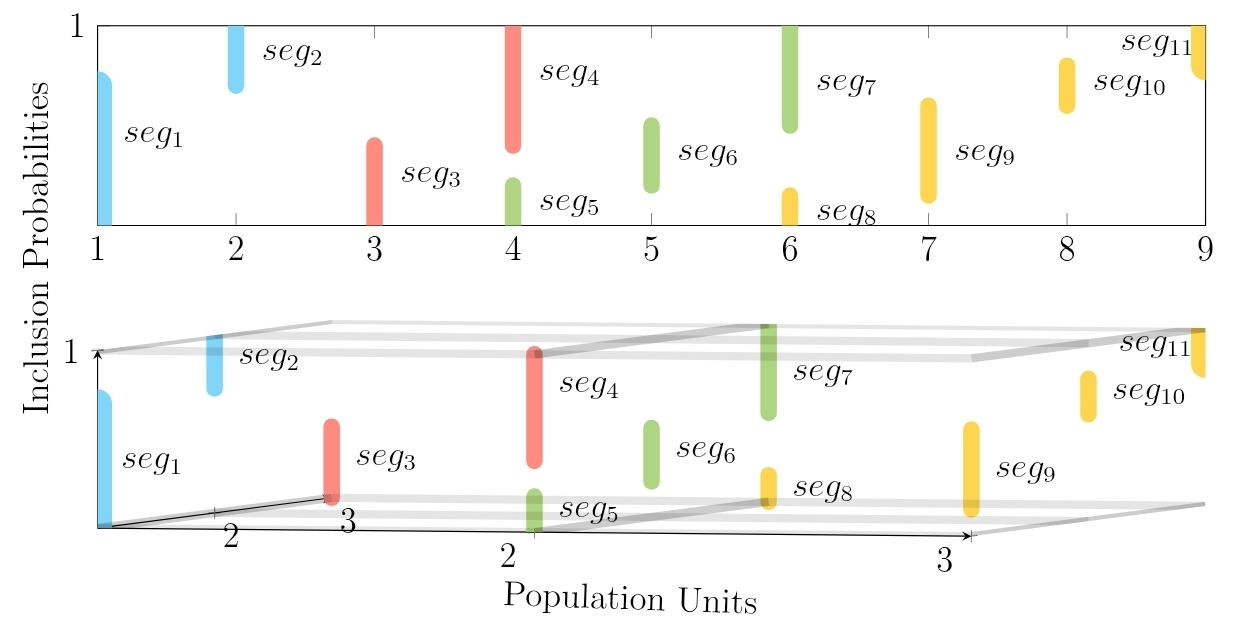}

    \caption{Visualization of the GFS design for one-dimensional (top) and two-dimensional (bottom) populations, as described in Example~\ref{PopExam1}. Vertical bars, potentially segmented by $seg_1,seg_2,\dots,seg_{11}$, represent each unit’s inclusion probability (UP) in lexicographic order. Colors indicate groups of units whose total UP equals~1, ensuring that a single uniform draw selects exactly one unit from each color group, thereby producing a fixed-size design.}
    \label{fig:GFSMadow}
\end{figure}

As two illustrative examples for $D=1$ and $D=2$, consider Figure \ref{fig:GFSMadow} with the following inclusion probabilities: {\small \begin{equation}\label{PopExam1} \pi_1 = 0.7,\pi_2 = 0.3,\pi_3 = 0.4,\pi_4 = 0.8,\pi_5 = 0.3,;\pi_6 = 0.65,\pi_7 = 0.45,\pi_8 = 0.2,\pi_9 = 0.2. \end{equation}} For the one-dimensional case, the population coordinate set is

$$\bm{C} = \{\bm{c}_1 = 1,\; \bm{c}_2 = 2,\; \bm{c}_3 = 3,\; \bm{c}_4 = 4,\; \dots,\; \bm{c}_9 = 9\},$$

$$\bm{C} = \{\bm{c}_1 = (1,1),\; \bm{c}_2 = (1,2),\; \bm{c}_3 = (1,3),\; \bm{c}_4 = (2,1),\; \dots,\; \bm{c}_9 = (3,3)\}.$$

In constructing the bars, we group those that collectively span the probability axis from \(0\) to \(1\) into a single stack. Within each stack, every bar requires at most two segments (indicated by $seg$ in Figure \ref{fig:GFSMadow}), splitting only when the cumulative probability reaches $1$ and wraps to $0$. By contrast, \citet{panahbehagh2025geometric} proposed an optional refinement that increases design entropy by subdividing bars into smaller pieces and rearranging them while preserving fixed sample size. Although this refinement could be incorporated here to further diversify the design, we do not pursue it in the present exposition.

The sampling step begins by drawing a random horizontal line across the ordered bars—this line represents a uniform random threshold within \([0,1]\). Units whose bars intersect this line are selected into the sample. Conceptually, units with bars aligned at similar heights have a synchronized chance of being selected together. This geometric mechanism naturally preserves first-order inclusion probabilities while introducing spatial coordination through the bar arrangement.

Arranging the bars to implement a sampling design affords precise control over the selection mechanism. By strategically choosing their placement and order, one can directly modulate the spatial spread and distribution of selected units, tailoring the design to specific objectives. This enables a form of “inverse engineering,” wherein desired sample properties guide the bar configuration: prescribing the intended allocation of selections across the population yields arrangements that achieve these targets while preserving the inherent randomness of the procedure.

The framework is highly flexible, supporting designs that meet statistical requirements and practical constraints simultaneously—for example, enforcing uniform spatial coverage or adjusting selection probabilities for designated units. Through careful configuration, it is possible to balance methodological rigor with operational needs, maintaining the robustness of probabilistic selection throughout.

\section{n-Means Spatial Sampling} \label{Sec:n-means}
In this section, we introduce \emph{$n$-Means Spatial Sampling}, a design for drawing well-spread samples from a finite population with prescribed inclusion probabilities. The design uses nested clustering to capture both global and local spatial structure. Its core rule is twofold; units from the same cluster are not selected together, while units occupying the same relative position across different clusters are selected jointly. Leveraging the GFS framework, we translate this rule into a sampling design by constructing a total order of the population units induced by the nested clustering.

The procedure first uses the $n$-Means UP-balanced clustering to shape spatially compact clusters $\{U_1,\ldots,U_n\}$ such that $\sum_{\ell \in U_i}\pi_\ell=1$ for every $i\in\{1,\ldots,n\}$. Within each cluster $U_i$, units are further partitioned into $m$ zones $\{U_{i,1},\ldots,U_{i,m}\}$ satisfying $\sum_{\ell\in U_{i,j}}\pi_\ell=1/m$ for each $j\in\{1,\ldots,m\}$, obtained by a second application of $n$-Means UP-balanced clustering targeted to the within-cluster total $1/m$. The $n$-Means UP-balanced clustering provide a global spatial structure for the population; units within each cluster \(U_i\) are geographically similar, so co-selecting them would degrade spread and is therefore avoided. However, cluster-level control alone may be insufficient, because units from two neighboring clusters can lie close to a shared boundary; selecting such pairs together would again reduce spread. To regulate this local geometry, we rely on the within-cluster zones. For examples of within‐cluster zones, see the shaded backgrounds labeled 1--4 in the first two rows of Figure~\ref{fig:three_popu_clusters}. Within the GFS framework, the guiding principle is to order the UP bars so that zones occupying the same relative position across clusters—thus inducing a well-spread zoning—begin at the same height and are processed in parallel during selection. Under this alignment, selecting units from equally labeled zones naturally disperses the sample across the population, while boundary-near units from adjacent clusters typically fall into different zones and are not co-selected. This yields samples that are well spread at both global and local levels.

\begin{figure}[!htbp]
    \centering
    \includegraphics[width=\linewidth]{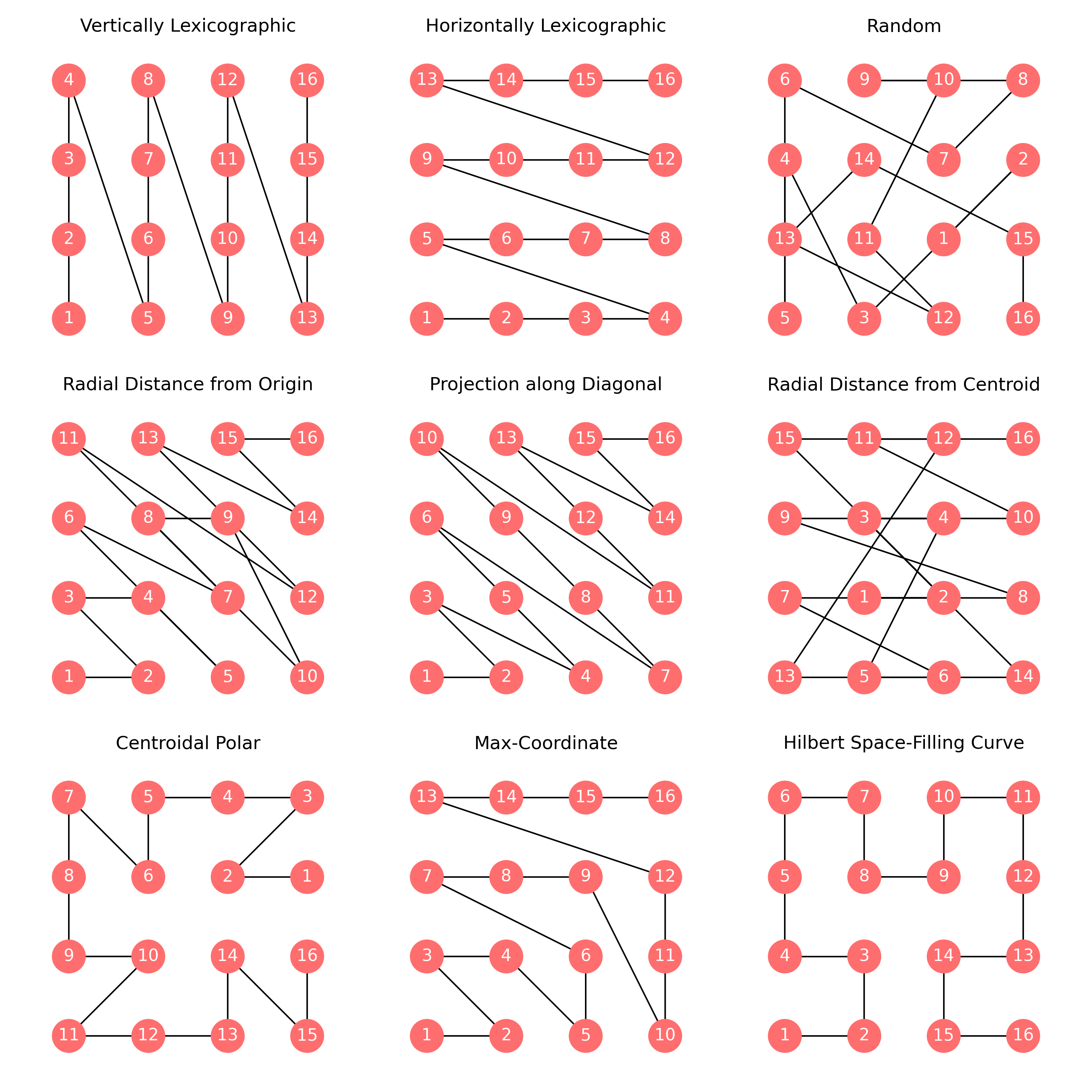}
    \caption{Illustrative ordering rules applied to spatial representatives—(i) zone centroids within a cluster or (ii) unit coordinates within a zone. In each panel, points are connected in ascending rank, visualizing the induced total order.}
    \label{fig:zones_order}
\end{figure}

To identify which zones occupy the same relative positions across clusters, we assign an order to the zones within each cluster via a ranking map \(\psi_1:\{1,\ldots,m\}\to\{1,\ldots,m\}\). The same ordering rule must be applied in every cluster to make ranks comparable. Figure~\ref{fig:zones_order} illustrates several such rules, including \emph{Horizontally Lexicographic}, \emph{Vertically Lexicographic}, \emph{Random}, \emph{Radial Distance from Origin}, \emph{Projection along Diagonal}, \emph{Radial Distance from Centroid}, \emph{Centroidal Polar}, \emph{Max-Coordinate}, and \emph{Hilbert Space-Filling Curve}. After applying the common ranking \(\psi_1\) in every cluster, zones receiving the same rank \(j\) are taken to define the same relative positioning zones across clusters. We list the zones of cluster \(i\) in this order as \(U_{i,(1)},U_{i,(2)},\ldots,U_{i,(m)}\).

An analogous ordering is imposed on units within each zone. For zone \(U_{i,(j)}\) and coordinates \(\mathbf{c}_\ell\in\mathbb{R}^2\) for \(\ell\in U_{i,(j)}\), define a ranking map \(\psi_2:U_{i,(j)}\to\{1,\ldots,|U_{i,(j)}|\}\) by ordering the units according to a common within-zone rule (e.g., any of the ordering schemes in Figure~\ref{fig:zones_order}) and letting \(\psi_2(\ell)\) be the position of unit \(\ell\) in that order. Using the same rule in every zone ensures comparable ranks. With this convention, units that share the same zone (by \(\psi_1\)) and the same within-zone rank (by \(\psi_2\)) are treated as having the same relative positioning and are scheduled to co-occur in samples, while units with different ranks are separated, enhancing local spread.

We now assemble the total order by traversing clusters along the fixed path supplied by \(n\)-Means UP-balanced clustering, and, within each cluster, concatenating zones and within-zone ranks with border adjustments. Denote the clusters in path order by \(U_{(1)},\ldots,U_{(n)}\). For each consecutive pair on this path, there is at most one border unit whose inclusion probability is split across the two clusters. In cluster \(U_{(i)}\), we list (1) the border unit with \(U_{(i-1)}\) first, if it exists; (2) the units of zones \(U_{(i),(1)},\ldots,U_{(i),(m)}\) in increasing within-zone rank; and (3) the border unit with \(U_{(i+1)}\) last, if it exists. If the same unit appears in consecutive positions in this construction (as may occur for border units), we collapse these into a single occurrence, ensuring that each population unit appears exactly once in the final order. Therefore, concatenating these within-cluster sequences in path order yields a unique total order \(\Psi\), which encodes all the intended global and local structure in a way such that the GFS framework could interpret it correctly.


The total order \(\Psi\) is tailored to the GFS construction. Because each cluster has total probability \(1\), every cluster forms exactly one complete stack, so precisely one unit from each cluster is selected in any sample—ensuring that units within the same cluster do not co-occur. Border units behave consistently with this scheme; a border unit is the one that splits across adjacent stacks, with its remainder initiating the next stack. Zones likewise align naturally. Since each zone carries probability \(1/m\) and zones of a cluster are placed contiguously in \(\Psi\), the cluster’s stack on \([0,1]\) is partitioned into \(m\) consecutive subintervals, each assigned to one zone. Consequently, zones with the same rank across clusters occupy the same subinterval of \([0,1]\); all samples arising from a given subinterval therefore select units from the same relative positioning zones in every cluster, achieving the intended cross-cluster co-occurrence pattern.

Providing the total order \(\Psi\) to the GFS framework completes the procedure and yields the \(n\)-Means Spatial Sampling design with the desired properties. Importantly, \(n\)-Means Spatial Sampling defines not a single design but a \emph{family} parameterized by the number of zones \(m\) per cluster and by the ranking functionals used to order zones and units. This structure offers flexibility within a systematic plan; fixing \(n\) clusters guarantees a baseline level of spread, while alternative within-cluster orderings induce distinct designs. At one extreme, choosing \(m=1\) (one zone per cluster) and a random within-zone ordering produces a high-entropy design within this family. At the other, adopting a larger \(m\) together with strict, geometry-aware ranking rules reduces entropy and strengthens spatial spread. This tunable trade-off between entropy and spread is valuable in practice, allowing the analyst to calibrate the design to application-specific priorities and constraints while preserving the prescribed first-order inclusion probabilities.

Algorithm~\ref{alg:nmeans-ss} summarizes the construction of \(n\)-Means Spatial Sampling.  It inputs spatial coordinates and inclusion probabilities, builds $n$-Means UP-balanced clustering and within-cluster zones, derives a total order \(\Psi\) consistent with the GFS framework, and finally draws a fixed-size, well-spread sample of size \(n\).

\subsection{Intelligent $n$-Means Spatial Sampling}
The $n$-Means Spatial Sampling leverages GFS to represent the population with $n$-Means UP-balanced clusters, zones, and simple ordering rules. This representation yields a \emph{family of valid designs}: by choosing the number of zones and the ordering of zones and units, we obtain many feasible options that all respect inclusion probabilities and sample size. The resulting flexibility makes it natural to \emph{search} for a design that best serves our goal—spreading the sample over the population. In this section, we make that search \emph{intelligent}: we use a greedy best-first procedure that proposes small, structure-preserving edits to a seed design, scores each candidate with a spreadness index, keeps the most promising versions, and explores nearby alternatives. Because edits act only on orderings (and, when allowed, the zone count), feasibility is preserved, evaluation is fast, and the approach adapts to the specific population at hand.

The algorithm consists of the following key stages:
\begin{itemize}
\item \textbf{Seeding:} Construct one or more plausible starting designs and score each with a chosen spreadness index. In the GFS framework, a design is fully specified by (i) the partition of the population into $n$-Means UP-balanced clusters, (ii) the number of zones per cluster, and (iii) the ordering rules applied to zones and to units within zones. Because these elements are native to GFS, every seed automatically satisfies inclusion probabilities and sample size constraints. All seeds are inserted into a max-priority queue keyed by their spread score.
\item \textbf{Selection:} Extract the top-ranked design from the queue to serve as the current parent. This concentrates computation on the most promising region of the design space while avoiding redundant expansion of the same candidate.
\item \textbf{Local variation:} Generate a small batch of child designs by applying limited, structure-preserving edits to the parent. Leveraging GFS’s graphical representation (UP bars, zones, and ordering primitives), edits are simple and safe: re-rank zones within selected clusters (e.g., align or permute equally labeled zones across clusters), adjust within-zone unit orders using predefined rules (e.g., radial, sweep, serpentine, space-filling). These operators modify only the ordering logic, so feasibility and design-based validity are maintained.
\item \textbf{Evaluation:} Score each child with the same spreadness index (e.g., DI, MI, VI, BI) and insert all children into the priority queue so that higher-scoring candidates rise to the top. The GFS data structures allow re-use of cluster/zone information, keeping evaluation costs low.
\item \textbf{Best-tracking:} If a child exceeds the best score observed so far, update the incumbent design and record the sequence of edits that produced it. This provenance makes improvements transparent and reproducible.
\item \textbf{Iteration or stop:} Repeat selection–variation–evaluation until an iteration budget is reached or a target spread level is achieved, then return the best design encountered. Because only local ordering changes are applied within GFS, the method is \emph{anytime}: it yields a valid, progressively improved design at every iteration and remains compatible with design-based inference throughout.
\end{itemize}
Further implementation details appear in Algorithm~\ref{alg:gbfs}.

 In the simulation study, we show that this approach reliably upgrades reasonable seeds to designs with substantially higher spread, both globally and within clusters.

\begin{algorithm}[!htbp]
\caption{\(n\)-Means Spatial Sampling}
\label{alg:nmeans-ss}
\begin{algorithmic}[1]
\Require Population \(U=\{1,\dots,N\}\) with coordinates \(\{\bm{c}_\ell\in\mathbb{R}^2\}_{\ell\in U}\); inclusion probabilities \(\bm{\pi}=(\pi_\ell)_{\ell\in U}\) with \(\sum_{\ell}\pi_\ell=n\).
\Require Number of zones per cluster \(m\ge 1\); zone-ranking function \(\psi_1\); within-zone ranking function \(\psi_2\).
\Ensure A fixed-size, well-spread sample \(S \in \mathcal{S}\) with \(|S|=n\).
\vspace{3pt}
\Statex \textbf{UP-Balanced clustering and path}
\State Run \textbf{\(n\)-Means UP-balanced clustering} to obtain clusters \(\{U_i\}_{i=1}^n\) with \(\sum_{\ell\in U_i}\pi_\ell=1\) and a fixed path order \(U_{(1)},\ldots,U_{(n)}\).\label{line:nmss-clusters}
\State For each consecutive pair \(\big(U_{(i)},U_{(i+1)}\big)\), identify at most one border unit \(b_{(i)}\) whose probability is split across the two clusters (if present).\label{line:nmss-borders}
\vspace{3pt}
\Statex \textbf{Within-cluster zoning and rankings}
\State For each \(i\), run \textbf{\(n\)-Means UP-balanced clustering} on \(U_i\) \emph{targeting \(m\) zones} with per-zone total \(1/m\), yielding \(\{U_{(i),1},\ldots,U_{(i),m}\}\).\label{line:nmss-zones}
\State Using \(\psi_1\) ranks zones as \(U_{(i),(1)},\ldots,U_{(i),(m)}\).\label{line:nmss-phi}
\State For each zone, using \(\psi_2\) to rank \(\ell\in U_{(i),(j)}\).\label{line:nmss-psi}
\vspace{3pt}
\Statex \textbf{Total order compatible with GFS}
\For{$i=1$ to $n$} \label{line:nmss-order-loop}
  \State Initialize \(\text{Stack}_{(i)}\gets [\,]\).
  \If{border with \(U_{(i-1)}\) exists} prepend \(b_{i-1}\) to \(\text{Stack}_{(i)}\). \label{line:nmss-front-border}
  \For{$j=1$ to $m$}
     \State Append units of \(U_{(i),(j)}\) to \(\text{Stack}_{(i)}\) in increasing \(\psi_2\). \label{line:nmss-append-zone}
  \EndFor
  \If{border with \(U_{(i+1)}\) exists} append \(b_i\) to the end of \(\text{Stack}_{(i)}\). \label{line:nmss-end-border}
  \EndIf
  \EndIf
\EndFor
\State Concatenate \(\text{Stack}_{(1)},\ldots,\text{Stack}_{(n)}\) to form a preliminary sequence; if the same unit appears in consecutive positions (due to border handling), keep a single occurrence. Denote the resulting total order by \(\Psi=(\Psi_1,\ldots,\Psi_N)\). \label{line:nmss-dedup}
\vspace{3pt}
\Statex \textbf{GFS sampling from the order}
\State Build \(n\) stacks on \([0,1]\) by sequentially placing bars of lengths \(\pi_{\Psi_1},\ldots,\pi_{\Psi_N}\); when a bar would cross 1, split it so the leading piece closes the current stack and place the remainder at 0 to start the next stack. \label{line:nmss-stacks}
\State Draw a single \(r\sim U(0,1)\); from each stack, select the unique unit whose bar intersects the vertical line at abscissa \(r\). \label{line:nmss-draw}
\State \Return The \(n\) selected units \(S\).
\end{algorithmic}
\end{algorithm}

\begin{algorithm}[!htbp]
\caption{An Intelligent \(n\)-Means Spatial Sampling}
\label{alg:gbfs}
\begin{algorithmic}[1]
\Require Population \(U=\{1,\dots,N\}\) with coordinates \(\{\bm{c}_\ell\}_{\ell\in U}\); inclusion probabilities \(\bm{\pi}=(\pi_\ell)_{\ell\in U}\).
\Require Set of initial designs \(\mathcal{D}esign_0=\{Design^{(1)},\dots,Design^{(q)}\}\) (each with fixed number of zones \(m\) and predefined zone and within-zone ordering rules).
\Require spreadness index \(\textsc{Spread}(\cdot)\) for scoring designs; iteration limit \(L\in\mathbb{N}\); optional target spread \(\tau\in\mathbb{R}\).
\Ensure A design \(Design^\star\) with high spread under \(\textsc{Spread}(\cdot)\).
\vspace{3pt}
\State Initialize a max-priority queue \(Q\) keyed by \(\textsc{Spread}(Design)\). \label{line:gbfs-initQ}
\For{each \(Design\in\mathcal{D}esign_0\)}
  \State Insert \(Design\) into \(Q\) with key \(\textsc{Spread}(Design)\). \label{line:gbfs-seed}
\EndFor
\State \(Design^\star \gets \arg\max_{Design\in\mathcal{D}esign_0}\textsc{Spread}(Design)\); \(\text{best} \gets \textsc{Spread}(Design^\star)\).
\State \(\ell \gets 0\). \Comment{iteration counter}
\vspace{3pt}
\While{\(\ell < L\) \textbf{and} \(Q\neq\emptyset\) \textbf{and} \((\tau\ \text{unset} \ \textbf{or}\ \text{best}<\tau)\)} \label{line:gbfs-while}
  \State \(Design_{\text{parent}} \gets \textsc{PopMax}(Q)\). \Comment{top of the priority queue} \label{line:gbfs-pop}
  \State Generate a set \(\mathcal{C}\) of child designs by copying \(Design_{\text{parent}}\) and applying one or more random modifications to: \label{line:gbfs-expand}
  \Statex \hspace{1.5em}\(\triangleright\) the ordering of zones in one or several clusters; and/or
  \Statex \hspace{1.5em}\(\triangleright\) the ordering of units in one or several zones.
  \For{each \(Design_{\text{child}}\in\mathcal{C}\)}
     \State Compute \(spread \gets \textsc{Spread}(Design_{\text{child}})\). \label{line:gbfs-score}
     \State Insert \(Design_{\text{child}}\) into \(Q\) with key \(spread\). \label{line:gbfs-push}
     \If{\(spread > \text{best}\)}
        \State \(Design^\star \gets Design_{\text{child}}\); \(\text{best} \gets spread\). \label{line:gbfs-updatebest}
     \EndIf
  \EndFor
  \State \(\ell \gets \ell + 1\).
\EndWhile
\State \Return \(Design^\star\).
\vspace{3pt}
\Statex \textbf{Notes:}
\begin{itemize}
\item Line~\ref{line:gbfs-score}: The spread score is computed by a chosen index from those introduced earlier (e.g., DI/MI/VI/BI).
\end{itemize}
\end{algorithmic}
\end{algorithm}

\section{Simulations}\label{Sec:Simulation}
In this section, we compare the proposed design and spreadness index with established spatially balanced sampling methods and existing indices, using a diverse collection of synthetic and real-world populations. To ensure transparency and reproducibility, all algorithms, population generators, and simulation scripts are implemented and publicly available in the \texttt{graphical-sampling} package \citep{graphical_sampling_2025}, accessible at \url{https://github.com/mehdimhb/graphical-sampling}. The simulation study is organized around three main components: the populations to which the designs are applied, the set of sampling designs included for comparison, and the indices used to evaluate the degree of spatial spread, as detailed below.

\begin{figure}[!htbp]
		\centering
\includegraphics[width=100mm]{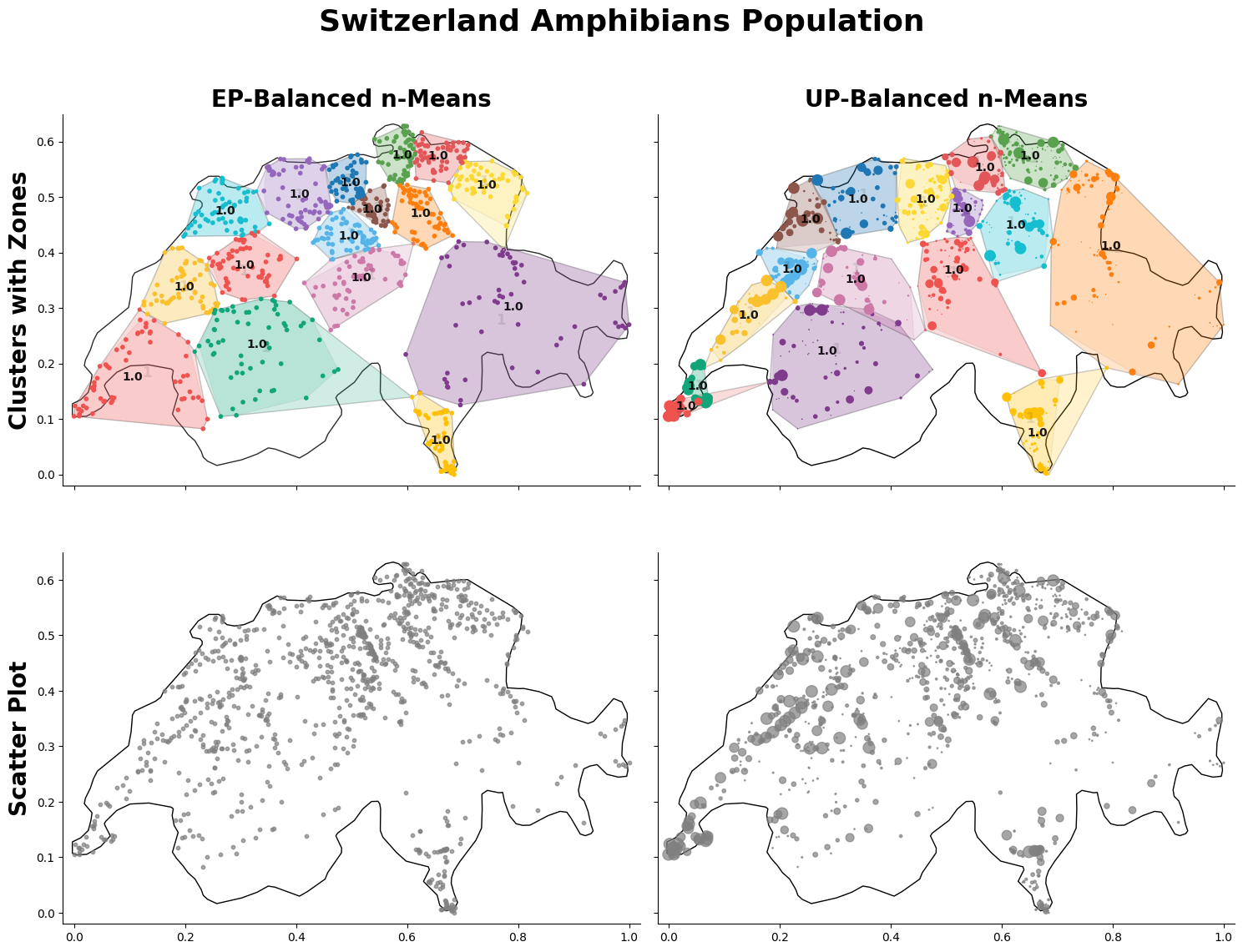} 
		\caption{ Amphibian observation sites in Switzerland under equal (EP, left) and unequal (UP, right) probabilities based on site area. Bottom row: scatter plots of sites; top row: $n$-Means UP-balanced clustering, one sample per cluster.} \label{fig:swiss_amphibian}
	\end{figure}

\begin{figure}[!htbp]
		\centering
\includegraphics[width=100mm]
{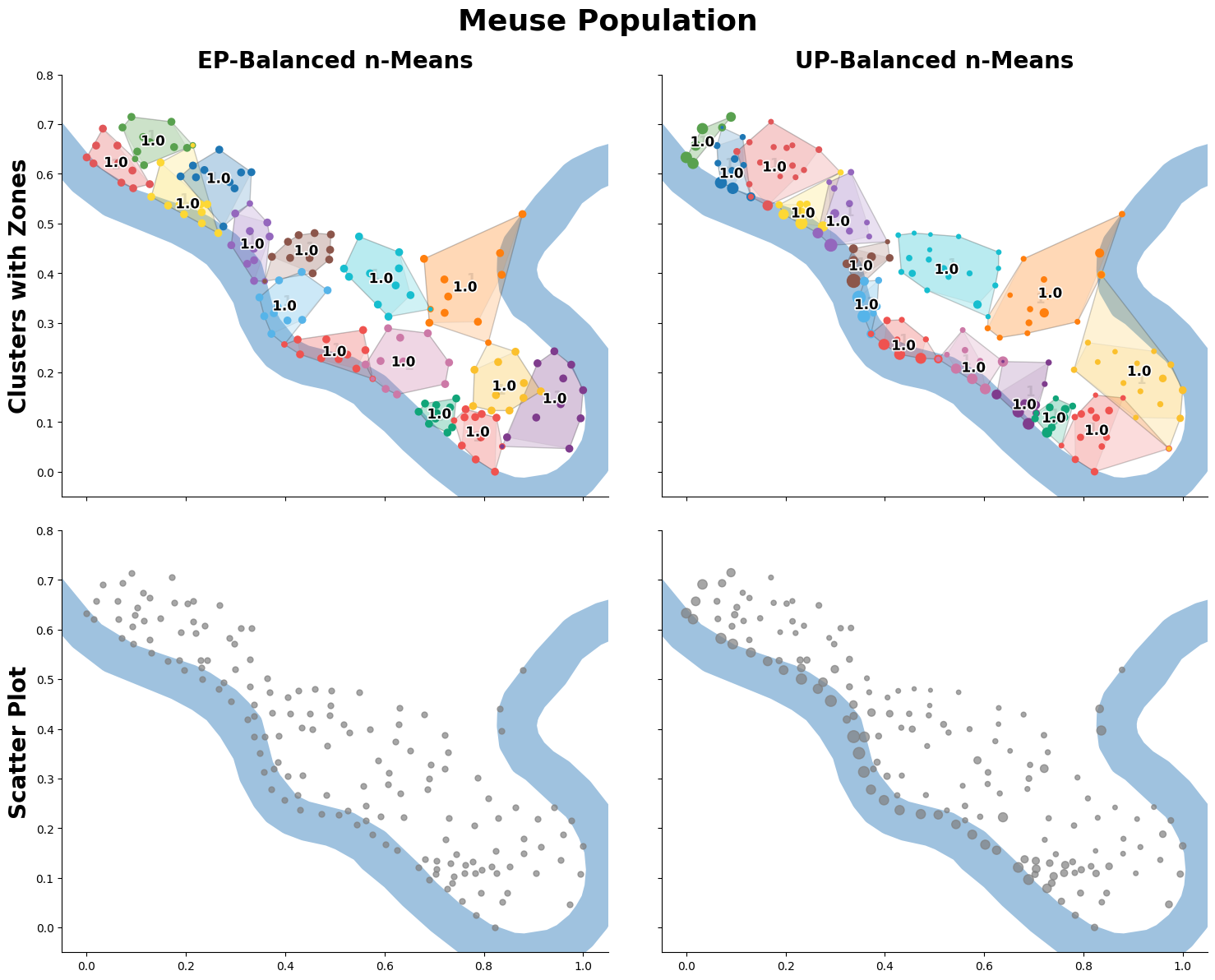} 
		\caption{ Meuse population showing sampling sites along the river under equal (EP, left) and copper-based unequal (UP, right) probabilities. Bottom row: scatter plots; top row: $n$-Means UP-balanced clustering, one sample per cluster.}\label{fig:Meuse_popu}
        \end{figure}

\begin{figure}[H]
    \centering
\includegraphics[width=100mm]{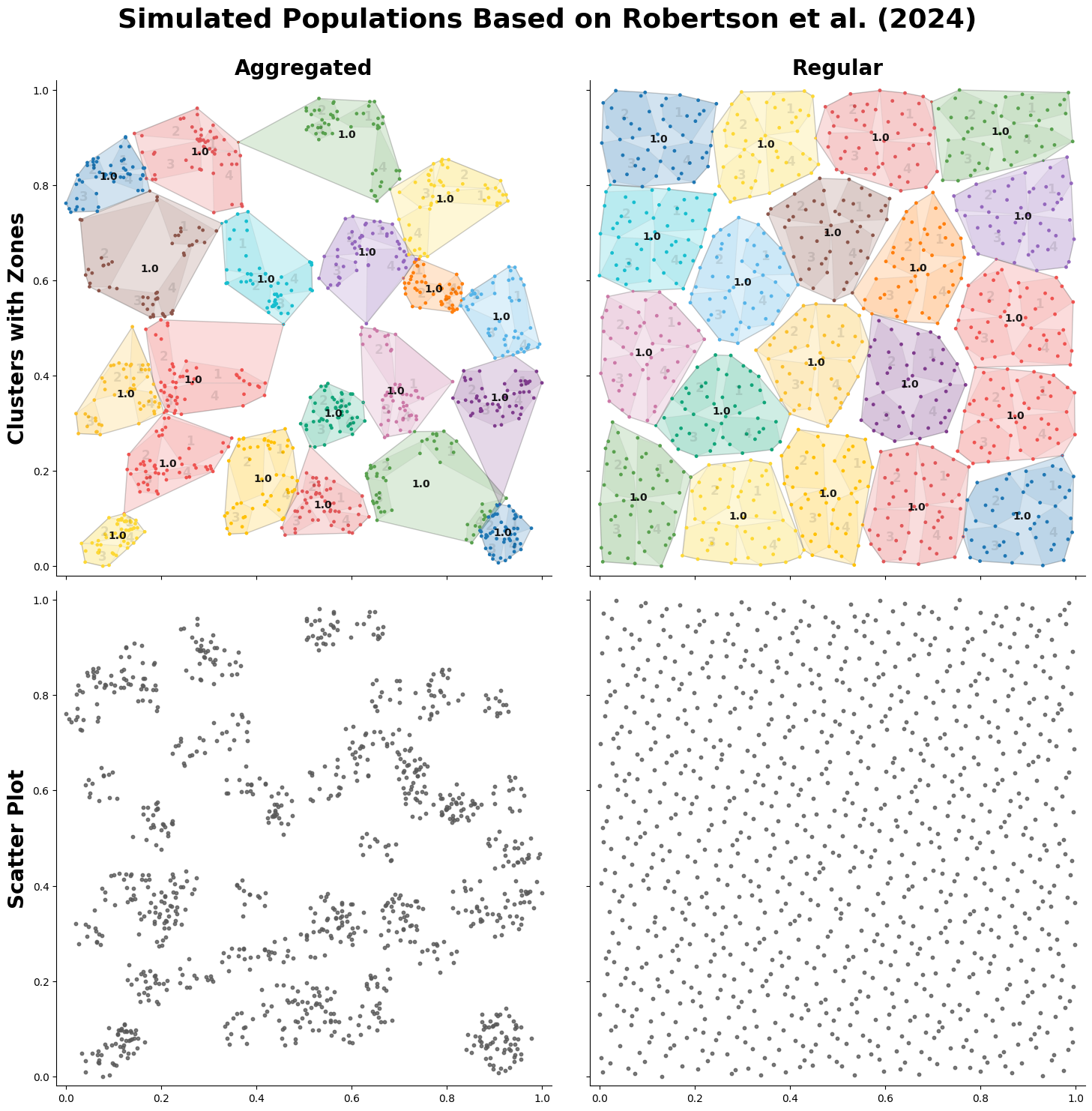}
\caption{Simulated populations from Robertson et al.~\cite{robertson2024}. Bottom row: scatter plots of aggregated (left) and regular (right) configurations; top row: $n$-Means UP-balanced clustering, one sample per cluster.}
    \label{fig:Robertson}
\end{figure}

\begin{itemize}
  
\item \textbf{Populations:}
\begin{itemize}

\item \textbf{synthetic Populations} (Figure~\ref{fig:three_popu_original}):\\
Three synthetic population types of size \(N=100\) were considered: gridded, random, and clustered configurations. Each population was evaluated under both equal-probability (EP) and unequal-probability (UP) scenarios. In the UP case, inclusion probabilities increase gradually from left to right, represented by larger point sizes for higher-probability locations. This setting provides a controlled framework to examine how the \(n\)-Means method adapts to both uniform and spatially varying inclusion-probability structures.

\item \textbf{Swiss Amphibians Population}\footnote{\url{https://www.karch.ch}} (Figure~\ref{fig:swiss_amphibian}):\\
The dataset, provided by the \textit{Centre de coordination pour la protection des amphibiens et des reptiles de Suisse} (karch), contains spatial coordinates for 959 observation sites across Switzerland along with site-level attributes. Geographic coordinates were used as spatial information, and the variable \emph{Area} (surface extent of each site) served as the auxiliary variable to define UP inclusion probabilities. Extremely large \emph{Area} values were truncated at 150 to prevent over-weighting.  

When applying one of the most computationally demanding spread designs, the WAV method, a runtime warning\footnote{\scriptsize\texttt{WARNING: Your population size is greater than 500; this could be quite time-consuming...}} indicated that simulations would be prohibitively slow for \(N=959\). To retain the spatial and auxiliary-variable structure while ensuring feasibility, we constructed a reduced population using conventional \(k\)-means clustering with \(k=100\). The observation nearest to each cluster centroid was retained, yielding a representative population of size \(N=100\).

\item \textbf{Meuse Population} (Figure~\ref{fig:Meuse_popu}):\\
The Meuse dataset, available in the \texttt{sp} package of R \citep{packagesp}, contains spatial coordinates and soil characteristics measured in the floodplain of the River Meuse near Stein, the Netherlands. The data include topsoil heavy-metal concentrations and several soil and landscape attributes, based on composite samples covering areas of approximately \(15 \times 15\) meters. In our analysis, spatial coordinates were used as location information, and copper concentration (\emph{Cu}) was employed as the auxiliary variable to define UP inclusion probabilities.

\item \textbf{Simulated Regular and Aggregated Populations} (Figure~\ref{fig:Robertson}):\\
To further evaluate performance of our desing, we used simulated populations introduced by \citet{robertson2024}. Two configurations were examined: (i) \emph{Regular populations}, generated as balanced acceptance samples \citep{Robertson2013} with \(N=1000\) points over the unit square; and (ii) \emph{Aggregated populations}, consisting of \(N=1027\) points generated via a conditional Neyman–Scott process with clustered spatial structure. The regular configuration emphasizes uniform spatial spread, whereas the aggregated configuration represents strongly clustered distributions, providing complementary benchmarks for assessing design robustness.
\end{itemize}

  \item \textbf{Sampling Designs:}
The following sampling designs were included in the simulations. Some were implemented directly using the authors’ available code, while others were faithfully reproduced based on the descriptions and results reported in their respective papers.
\begin{itemize}

  \item \textbf{WAV:}\\
  The Weakly Associated Vector method constructs spatially balanced samples by iteratively perturbing the inclusion-probability vector along directions weakly associated with spatial proximity, progressively fixing units while preserving first-order inclusion probabilities \citep{jauslin2020spatial}.
  
  \item \textbf{LP1:}\\
  The Local Pivotal Method (variant 1) applies pairwise updates to neighbouring units so that exactly one is retained in each step. This induces spatial repulsion and yields samples with strong local balance and reduced clustering \citep{gra:lun:sch:12, graf2016}.
  
  \item \textbf{SCP:}\\
  Spatially Correlated Poisson sampling introduces dependence among the selection indicators of nearby units, reducing the likelihood of jointly including close neighbours and improving the spatial spread of the resulting sample \citep{graf:11}.
  
  \item \textbf{GRT:}\\
  Generalized Random Tessellation Stratified sampling maps multi-dimensional populations to a one-dimensional ordered frame through recursive tessellation, from which systematic selection produces well-spread, spatially balanced samples \citep{Stev:Olse:spat:2004}.
  
  \item \textbf{HIP:}\\
  Halton Iterative Partitioning exploits the low-discrepancy properties of Halton sequences to recursively partition the spatial domain, generating quasi-random, uniformly distributed samples suitable for finite two-dimensional populations \citep{robertson2018halton}.
  
  \item \textbf{DAS:}\\
  Dynamic Assignment Sampling formulates the selection process as a sequence of linear assignment problems that iteratively optimize geometric spread while maintaining prescribed inclusion probabilities. It supports dynamic sample-size adjustment and general auxiliary structures \citep{robertson2024}.
    
\item \textbf{NMS} (Algorithm \ref{alg:nmeans-ss}):\\
$n$-Means Spatial Sampling partitions the population into $n$-means UP-balanced clusters and, within each cluster, into zones; a GFS step is then applied to achieve both global and local spatial spread. Each cluster has total inclusion probability $1$ and is subdivided into $m=4$ zones, each with total inclusion probability $0.25$ . In our simulations, we used the \emph{Centroidal Polar} ranking, as defined in Section~\ref{Sec:n-means}, both for ordering zones and for ordering units within zones. Comparative experiments indicated that the choice of ranking can materially affect NMS performance; among the alternatives we tested, \emph{Centroidal Polar} was generally the most efficient. A comprehensive analysis of ranking schemes is of independent interest; in this study, we standardize on \emph{Centroidal Polar} to focus on the main design features.

\item \textbf{GMS} (Algorithm \ref{alg:gbfs}):\\
An intelligent greedy search built on NMS refines the GFS ordering while preserving fixed sample size and first-order inclusion probabilities. Starting from one or more NMS-based initial designs (fixed , predefined orderings), the procedure iteratively proposes local reorderings of zones within clusters and of units within zones, evaluates candidates using a spreadness index, and adopts the best-scoring configuration at each step. In our simulations, MI served as the guiding index. The search terminates when a target spread is achieved or a prespecified computational budget is reached.

  \item \textbf{SRS:}\\
  Simple Random Sampling Without Replacement serves as the baseline equal-probability design. Although it does not target spatial balance, it provides a natural benchmark for evaluating the gains achieved by spread-oriented methods \citep{til:06}.
  
  \item \textbf{MAX:}\\
  Maximum-Entropy sampling selects samples that maximize entropy subject to given inclusion probabilities. While not explicitly spatial, it offers a design-based benchmark for assessing how spatially oriented algorithms enhance representativeness \citep{til:06}.
\end{itemize}

Cube-based and sequential balancing variants proposed by \citet{jauslin2022sequential} and \citet{Panahbehagh2023GSS} were not included. Although these approaches have potential to promote spatial balance through auxiliary-variable control, they have not yet been extended to operate directly on spatial coordinates or to incorporate geometric distances into their balancing mechanism.

\item \textbf{Indices:}
\begin{itemize}
\item \textbf{Moran Index  (MI):} Assesses spatial autocorrelation via a spatial-weights matrix; negative values indicate dispersion, positive values indicate clustering, and values near zero indicate spatial randomness \citep{moran1950notes,til:dic:esp:giu:18}.
\item \textbf{Voronoi Index (VI):} Measures deviation between observed and expected inclusion probabilities across Voronoi cells; lower values indicate greater spatial balance, while higher values indicate clustering or uneven coverage \citep{Stev:Olse:spat:2004}.
\item \textbf{Balanced Voronoi Index (BI):} Extends VI by incorporating auxiliary-variable balance within cells; lower values indicate both spatial and auxiliary balance, whereas higher values signal imbalance across one or both dimensions \citep{prentius2024spatial}.
\item \textbf{Density Disparity Index (DI):} Quantifies spread through a translation-invariant, signed comparison under cluster-wise translations (balanced -means); values near zero indicate well-spread samples, negative values reflect over-dispersion, and positive values reflect under-dispersion.

\end{itemize}

\end{itemize}

Using the selected populations, designs, and indices, we conducted extensive simulation experiments. Practical constraints limited direct replication of some competitors: WAV became computationally prohibitive for larger \(N\), and the available DAS implementation is in \texttt{Matlab}, which we could not feasibly integrate into our pipeline. Accordingly, we adopted the following protocol. When author-supplied generators were available, we regenerated their populations and applied our designs and indices under identical settings. When faithful reimplementation was not practical, we reported the competing method’s results as published in the respective sources on the same (or author-provided) populations and metrics, explicitly indicating their provenance. This approach preserves comparability while maintaining transparency regarding which results are reproduced and which are drawn from prior work.

It is important to emphasize that the focus of this paper is solely on assessing the efficiency of sampling designs in terms of their ability to spread the sample. To avoid diverting attention from this objective, we did not consider efficiency with respect to parameter estimation (such as population totals) or issues related to variance estimation under these designs.  

Preliminary comparisons indicated that WAV and LP1 consistently produced more efficient spreading than most other existing methods. Based on this, the simulations proceeded as follows. First, for the three synthetic populations and the Swiss amphibian population, we compared our proposed designs---NMS and GMS---with WAV, LP1, SRS, and MAX. For the Meuse population, we relied on the results of \citet{jauslin2020spatial}, and in the case of $n=15$, we extended the comparison of NMS and GMS to include WAV, LP1, SCP, GRT, MAX, and SRS. Finally, to evaluate our methods against DAS, we incorporated the simulation results of \citet{robertson2024}, where we compared GMS with GRT, HIP, LP1, and DAS.

\subsection{Results for Clustered, Gridded, Random, and Swiss Populations}
\begin{figure}[!htbp]
		\centering
\includegraphics[width=150mm]{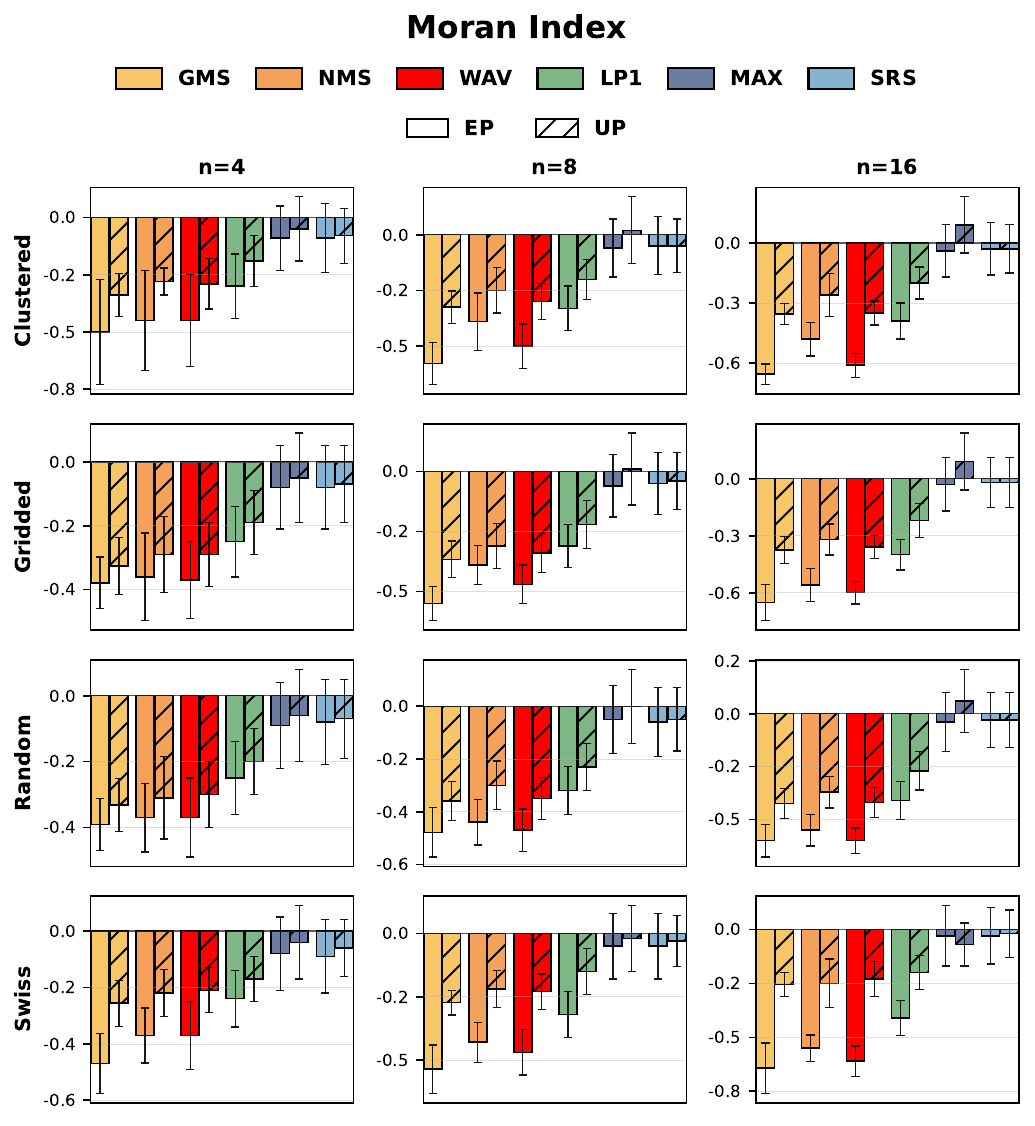} 
		\caption{Results from Monte Carlo simulations under equal (EP) and unequal (UP) probability sampling schemes. Boxplots display the distribution of the Moran index obtained from different designs applied to three simulated populations (Clustered, Gridded, Random) and the observed Swiss population, with sample sizes $n=4,8,16$.}
\label{fig:Moran_Index}
	\end{figure}

\begin{figure}[!htbp]
		\centering
\includegraphics[width=150mm]{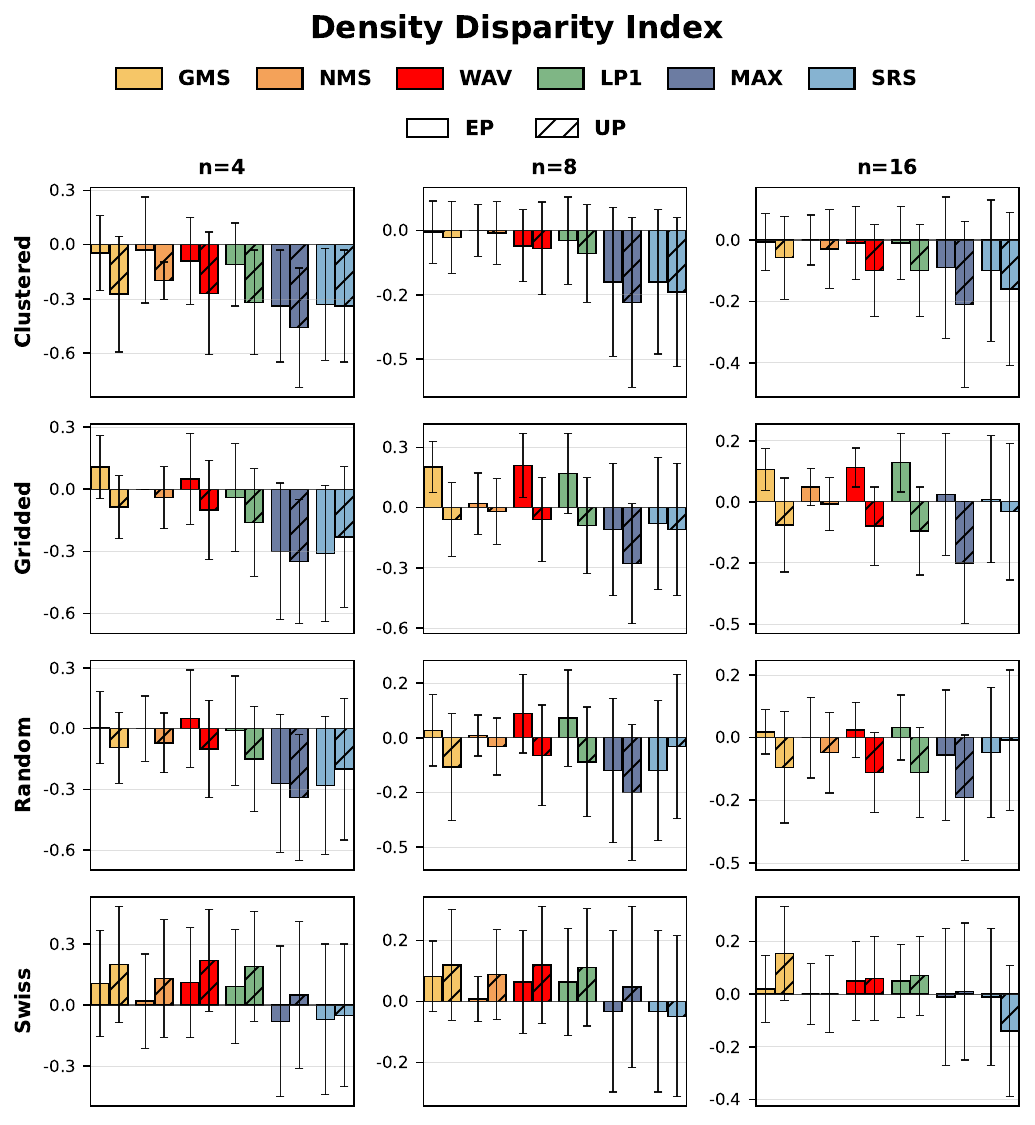} 
		\caption{Results from Monte Carlo simulations under  equal (EP) and unequal (UP) probability sampling schemes. Boxplots display the distribution of the Density Disparity Index obtained from different designs applied to three simulated populations (Clustered, Gridded, Random) and the observed Swiss population, with sample sizes $n=4,8,16$.}
\label{fig:Density_Disparity_Index}
	\end{figure}

\begin{figure}[!htbp]
		\centering
\includegraphics[width=150mm]{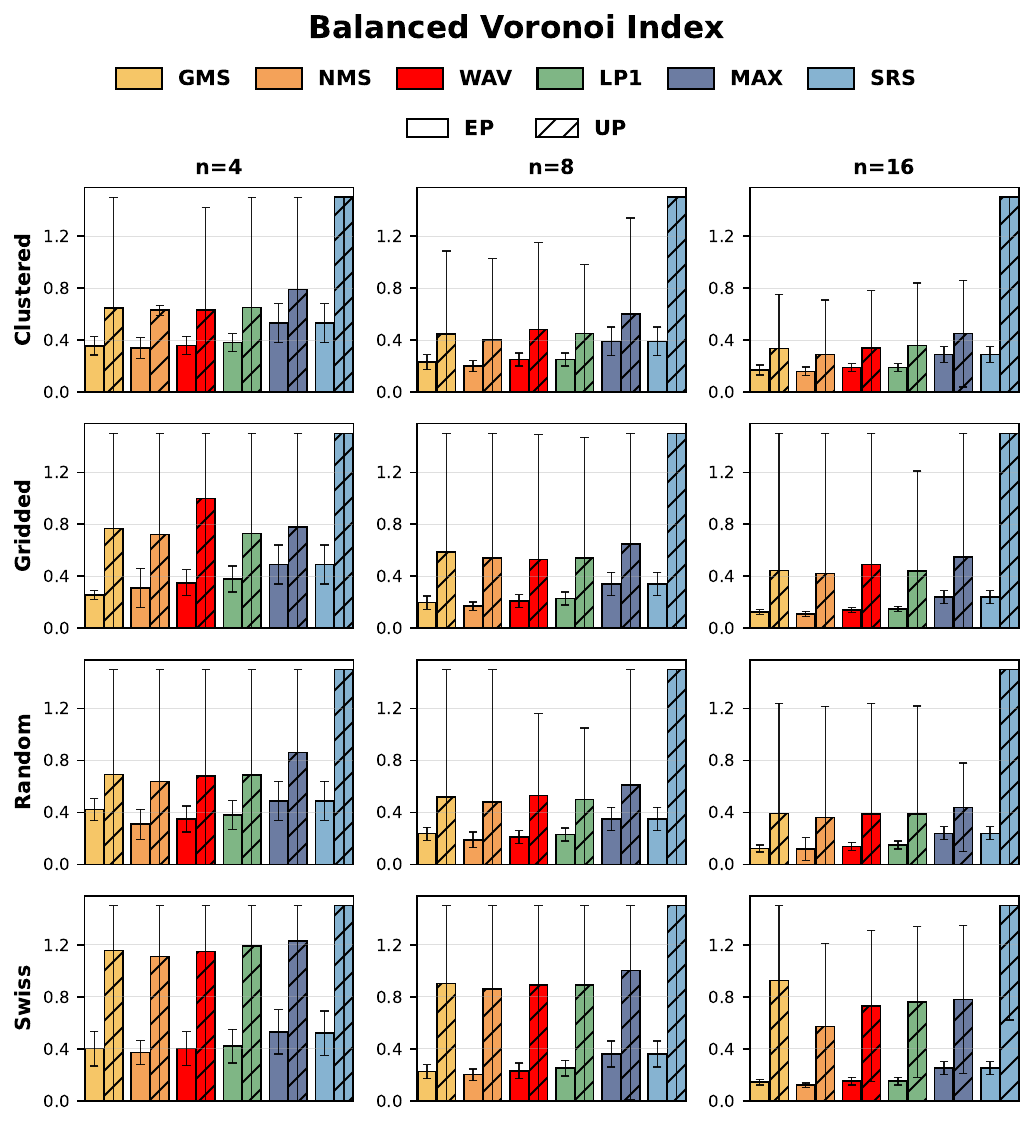} 
		\caption{Results from Monte Carlo simulations under  equal (EP) and unequal (UP) probability sampling schemes. Boxplots display the distribution of the Balanced Voronoi Index obtained from different designs applied to three simulated populations (Clustered, Gridded, Random) and the observed Swiss population, with sample sizes $n=4,8,16$.}
 \label{fig:Balanced_Voronoi_Index}
	\end{figure}

The results of the Monte Carlo simulations are presented in Figures~\ref{fig:Moran_Index}, \ref{fig:Density_Disparity_Index}, and \ref{fig:Balanced_Voronoi_Index}, corresponding to MI, DI, and BI, respectively. As the BI represents a refined and extended version of the VI, incorporating auxiliary-variable balance, we report only BI to represent the family of Voronoi-based indices and avoid redundancy in presentation.

Across the three indices considered---MI, DI, and BI---the relative performance of the designs was broadly consistent with expectations, while also revealing several noteworthy findings. WAV and LP1 achieved strong spatial dispersion across all indices, with WAV consistently—and in most cases significantly—outperforming LP1. On BI and DI, NMS surpassed WAV, whereas on MI, WAV outperformed NMS in most settings. Motivated by this contrast, GMS was implemented with MI as the optimization criterion; as intended, GMS surpassed WAV on MI across all populations and sample sizes, with gains particularly pronounced for the Swiss data, while remaining competitive on BI and DI (generally comparable to WAV and LP1).

SRS and MAX, by contrast, demonstrated clear limitations in spreading samples, as anticipated. Their weaknesses were particularly evident under DI, which revealed a much broader range of outcomes (\(-0.60\) to \(0.30\)) compared with MI (approximately \(-0.15\) to \(0.15\)). This extended range highlights the sensitivity of the DI in detecting differences among designs and its ability to capture the long-tailed behavior of weakly spreading methods. Although SRS is not ideal under unequal probabilities, it was retained in the simulations to preserve comparability across plots.

For BI, occasional extreme values exceeding one were observed, which complicated interpretation and distorted the figures; accordingly, the plots were clipped at 1.22. 

Overall, the DI demonstrated particular interpretability, with values around zero indicating well-spread samples, increasingly negative values reflecting stronger dispersion, and positive values indicating concentration. This directional property makes DI a valuable complement to existing indices for assessing spatial spread.


 \subsection{Results for the Meuse Population}
 \begin{figure}[!htbp]
		\centering
\includegraphics[width=100mm]{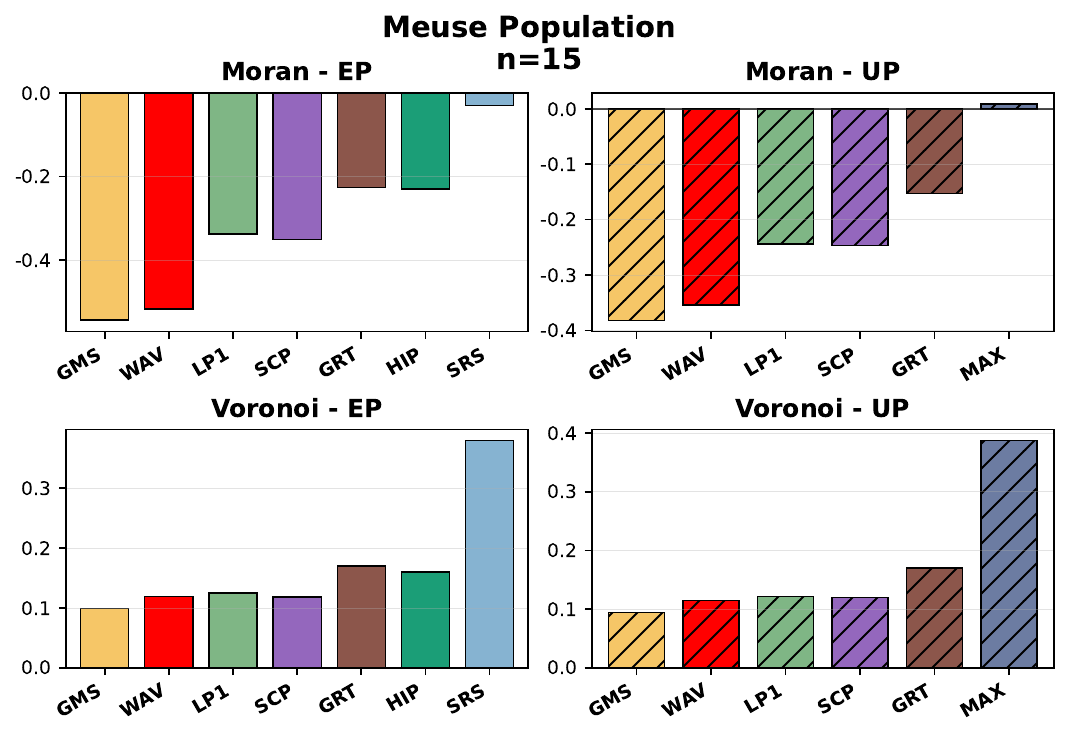} 
		\caption{Comparison of spatial spreading indices for the Meuse population with $n=15$. Results are shown under both EP and UP settings using Moran and the Voronoi Index. The proposed methods (GMS) are evaluated against existing spread-oriented designs (WAV, LP1, SCP, GRT, HIP) as well as baseline methods (SRS, MAX). } \label{fig:meuse-result}
	\end{figure}
The results for the Meuse population with $n=15$ clearly demonstrate the superior spreading performance of GMS relative to all competing methods. Under both EP and UP settings, and across both indices considered (Moran and the Voronoi index), GMS consistently achieves values closer to the optimal benchmark. This indicates that GMS produces samples that are more dispersed and spatially balanced than those obtained from established designs.  

The advantage of GMS is particularly pronounced when evaluated under the UP setting. In both Moran and Voronoi-based measures, the separation between GMS and the next-best competitors (notably WAV and LP1) is considerably larger in UP compared to EP. This highlights the robustness of GMS in handling heterogeneity induced by unequal inclusion probabilities, a common challenge in practical survey scenarios.

SRS and MAX serve here as natural baselines rather than true competitors. While other methods such as SCP, GRT, HIP, and WAV demonstrate improvements relative to these baselines, none achieve dominance over GMS. This consistent superiority suggests that the geometric construction underlying GMS provides a more stable and generalizable framework for spreading, ensuring reliable performance across both EP and UP settings.  

\subsection{Results for Simulated Regular and Aggregated Populations}
The results reported in Table~\ref{tab:compare2024} are based on the simulated regular and aggregated populations introduced by \citet{robertson2024}. For comparability, we reproduce the outcomes of GRT, HIP, LP1, and DAS from their study and supplement them with the performance of our proposed GMS design. Two indices are reported: the Voronoi index, where lower values indicate better spatial balance, and Moran, where more negative values indicate stronger dispersion.  

Across both regular and aggregated populations, GMS consistently achieves Moran values that are substantially more negative than those of all other designs. This indicates that GMS produces samples with a markedly stronger degree of spatial dispersion. The superiority of GMS is maintained across all sample sizes, from $n=20$ to $n=200$, and becomes even more pronounced as the sample size increases, where the gap between GMS and competing methods steadily widens.  

In terms of the Voronoi index, GMS also dominates all competitors. For both population structures and across all sample sizes, GMS yields the lowest Voronoi values, indicating a level of spatial balance that is unmatched by GRT, HIP, LP1, or DAS. These differences are not marginal.

Taken together, the results show that GMS achieves simultaneous improvements on both indices: it strongly outperforms in Moran by delivering greater dispersion, and at the same time it provides superior Voronoi values, reflecting balanced coverage. This dual advantage underscores the effectiveness of the geometric construction underlying GMS, which ensures robust performance across different spatial structures and sample sizes.
\begin{table}[ht]
\caption{Comparison of spatial spreading indices (Voronoi and Moran) for regular and aggregated simulated populations from \citet{robertson2024}. Results for GRT, HIP, LP1, and DAS are reproduced from their study, while GMS is newly added. GMS consistently attains more negative Moran and lower Voronoi-based index values than all competitors across all sample sizes, thereby dominating on both criteria.}
    \label{tab:compare2024}
\resizebox{\textwidth}{!}{
\begin{tabular}{lrrrrrrrrrr}

\toprule
$n$ & \multicolumn{2}{c}{GRT } & \multicolumn{2}{c}{HIP } & \multicolumn{2}{c}{LP1 } & \multicolumn{2}{c}{DAS } & \multicolumn{2}{c}{GMS} \\
    & Voronoi & Moran                   & Voronoi & Moran            & Voronoi & Moran            & Voronoi & Moran            & Voronoi & 
    Moran        \\
\midrule
\multicolumn{11}{l}{\textbf{Regular Population}} \\
20  & 0.38 & $-$0.09 & 0.25 & $-$0.13 & 0.24 & $-$0.16 & 0.15 & $-$0.24 & 0.08 & $-$0.25 \\
50  & 0.40 & $-$0.12 & 0.25 & $-$0.20 & 0.23 & $-$0.24 & 0.16 & $-$0.32 & 0.06 & $-$0.36 \\
100 & 0.44 & $-$0.16 & 0.30 & $-$0.24 & 0.24 & $-$0.32 & 0.20 & $-$0.38 & 0.06 & $-$0.42 \\
200 & 0.52 & $-$0.19 & 0.41 & $-$0.26 & 0.31 & $-$0.43 & 0.32 & $-$0.43 & 0.09 & $-$0.47 \\
\midrule
\multicolumn{11}{l}{\textbf{Aggregated Population}} \\
20  & 0.40 & $-$0.11 & 0.37 & $-$0.12 & 0.28 & $-$0.18 & 0.24 & $-$0.26 & 0.09 & $-$0.27 \\
50  & 0.44 & $-$0.14 & 0.42 & $-$0.17 & 0.29 & $-$0.26 & 0.27 & $-$0.34 & 0.12 & $-$0.38 \\
100 & 0.47 & $-$0.18 & 0.46 & $-$0.21 & 0.30 & $-$0.35 & 0.30 & $-$0.40 & 0.14 & $-$0.45 \\
200 & 0.52 & $-$0.22 & 0.51 & $-$0.25 & 0.33 & $-$0.46 & 0.37 & $-$0.45 & 0.17 & $-$0.48 \\
\bottomrule
\end{tabular}}

\end{table}

\section{Summary and Conclusion}\label{Sec:Conclusion}
This paper introduced a graphical and intelligent framework for assessing and enhancing spatial spread in finite-population sampling. We proposed the Density Disparity Index, a kernel-based measure comparing the original population density with its cluster-wise translated counterpart, to quantify spatial distortion induced by a sample. DI complements MI and BI by distinguishing local redistribution effects from global autocorrelation or imbalance across cells. In addition, we developed $n$-Means UP-balanced clustering to form probability-balanced, spatially compact groups that support representative single-unit sampling, and used this structure as the basis for a GFS optimized for spatial dispersion under design-based validity. An intelligent search mechanism was further integrated, yielding the GMS design, which adaptively refines GFS orderings to maximize spread while preserving inclusion probabilities. Monte Carlo experiments on vast range of populations demonstrate that GMS dominates all competing designs, achieving more negative Moran’s , close to zero DI, and lower or comparable BI values across sample sizes. The framework is practical—scaling efficiently through simple clustering and localized intelligent search—modular, allowing flexible choice of kernels and balancing targets, and inferentially coherent, preserving design-based validity.
Future work includes increasing the entropy of NMS by segmenting GFS bars into finer subcomponents, thereby enlarging the feasible neighborhood of reorderings while preserving fixed size and first-order inclusion probabilities. This higher-resolution search space would enable more intelligent optimization—e.g., genetic algorithms, artificial bee colony methods, and related metaheuristics—to explore diversified local and global permutations under design-based constraints.

\end{document}